%% file: Joint_Scheduling_and_PowerControl.tex
\pdfoutput=1 
\documentclass[conference]{IEEEtran}

\usepackage{cite}
\usepackage{graphicx}
\usepackage{amsmath,amsthm}
\usepackage{algpseudocode}
\usepackage{amssymb}
\usepackage{soul}			
\usepackage{algorithm}
\usepackage{algorithmicx}
\usepackage{algpseudocode}
\algdef{SE}[DOWHILE]{Do}{doWhile}{\algorithmicdo}[1]{\algorithmicwhile\ #1}%
\algrenewcommand\algorithmicrequire{\textbf{Input:}}
\algrenewcommand\algorithmicensure{\textbf{Output:}}

\newtheorem{lemma}{Lemma}
\makeatletter
\newcommand*{\rom}[1]{\expandafter\@slowromancap\romannumeral #1@}
\makeatother
\newcommand{\comments}[1]{}
\newcommand{\argmin}{\operatornamewithlimits{argmin}} 
\def\argmax{\mathop{\rm arg\,max}}
\usepackage{multirow}
\usepackage{relsize}     
\usepackage{verbatim}
\usepackage{blindtext}
\usepackage{amsmath}

\usepackage{makecell}
\usepackage{tabu,enumitem}
\usepackage{mathtools, cuted} 

\newcommand{\Ls}{\mathcal{L}}
\newcommand{\Ks}{\mathcal{K}}
\newcommand{\Hb}{\mathbf{H}}

\newcommand{\Rsbar}{\bar{\mathcal{R}}}
\newcommand{\timeslots}{\bar{\mathcal{T}}}
\newcommand{\Bs}{\mathcal{B}}

\newcommand{\acir}{A}
\newcommand{\acirf}{A_{f',f}}
\newcommand{\zeroM}{\mathbf{0}}

\DeclareMathOperator{\E}{E}			
\usepackage[usenames, dvipsnames]{color}
\usepackage{lipsum}
\usepackage[most]{tcolorbox}
\definecolor{block-gray}{gray}{0.85}
\newtcolorbox{myColoredquote}{colback=block-gray,
	boxrule=0pt,boxsep=0pt,breakable}
\definecolor{green1}{rgb}{0, 0.6, 0.3}   
\definecolor{blue1}{rgb}{0, 0, 0.8}   

\newcommand{\reviewerA}[1]{{\color{black}#1}}   
\newcommand{\reviewerB}[1]{{\color{black}#1}}    

\usepackage[pscoord]{eso-pic}

\input{patterntikz}		
\usepackage[utf8]{inputenc}
\usepackage{tikz}
\usetikzlibrary{shapes,arrows,chains,arrows,decorations.pathmorphing,backgrounds,positioning,fit,petri,calc,patterns}
\usetikzlibrary{matrix,positioning}
\usepackage{pgfplots}
\usepackage{caption,subcaption}
\usepgfplotslibrary{groupplots}
\pgfplotsset{compat=newest}

\usepackage[T1]{fontenc}
\usepackage{mathtools}
\usepackage{amssymb}

\usepackage[nocomma,short]{optidef}

\newcommand{\Pmax}{P^{\text{max}}}  
\newcommand{\Pinit}{P^{\text{init}}}

\newcommand{\gammaT}{\gamma^{\mathrm{T}}}
\newcommand{\gammaTbar}{\bar{\gamma}^{\mathrm{T}}}
\newcommand{\Pbar}{\bar{P}}

\newcommand{\Ct}{C^{\text{max}}}

\newcommand{\davg}{$d_{\text{avg}}$}

\newcommand{\xa}{{x}}
\newcommand{\xab}{{\mathbf{x}}}
\newcommand{\na}{{n}}
\newcommand{\nab}{{\mathbf{n}}}
\newcommand{\fa}{{f}}
\newcommand{\fab}{{\mathbf{f}}}
\newcommand{\Ns}{\mathcal{N}}
\newcommand{\Fs}{\mathcal{F}}
\newcommand{\Gs}{\mathcal{G}}
\newcommand{\Pb}{\mathbf{P}}
\newcommand{\Yb}{\mathbf{Y}}
\newcommand{\Ub}{\mathbf{U}}
\newcommand{\Xb}{\mathbf{X}}
\newcommand{\Zb}{\mathbf{Z}}

\newcommand{\Ft}{\tilde{F}}
\newcommand{\Pt}{\tilde{P}}

\newcommand{\figref}[1]{Fig.~\ref{#1}}
\newcommand{\round}[1]{\lfloor {#1} \rceil}
\usepackage{glossaries}
\makeglossaries
\newacronym{PA}{PA}{power amplifier}
\newacronym{LP}{LP}{linear programming}
\newacronym{BS}{BS}{base station}
\newacronym{VANETs}{VANETs}{vehicular ad hoc networks}

\IEEEoverridecommandlockouts
\begin{document}

\title{Scheduling and Power Control for V2V Broadcast Communications with Co-Channel and \\ Adjacent Channel Interference}
\author{%
	\vspace{-0.5cm}
	\IEEEauthorblockA{}{Anver~Hisham,  Erik~G.~Str\"{o}m, Fredrik~Br\"annstr\"om, and Li Yan }
	\IEEEauthorblockA{Department of Electrical Engineering, Chalmers University of Technology, Gothenburg, Sweden\\
		\emph{\{anver,~erik.strom,~fredrik.brannstrom, lyaa\}@chalmers.se}\\\vspace{-1.0cm}
		\thanks{The research was, in part, funded by the Swedish Governmental Agency for Innovation Systems (VINNOVA), FFI - Strategic Vehicle Research and Innovation, under Grant No. 2014-01387. This work has, in part, been performed in the framework of the H2020 project 5GCAR co-funded by the EU. The authors would like to acknowledge the contributions of their colleagues. The views expressed are those of the authors and do not necessarily represent the project. The consortium is not liable for any use that may be made of any of the information contained therein.}\\
	}
}

\maketitle \thispagestyle{empty}


\begin{abstract}
	\input{abstract.tex}

\end{abstract}

\input{sec01_introduction.tex}
\input{sec02_system_model.tex}
\input{sec02b_problemFormulation_ES.tex}

\input{sec03_contributions.tex}
\input{sec04_powerControl.tex}
\input{sec05_SimulationResults.tex}
\input{sec06_Conclusion.tex}
\bibliographystyle{IEEEtran}
\bibliography{references}
\end{document}

%% file: patterntikz.tex
\usepackage{tikz}
\usepackage{tikz-3dplot}
\usetikzlibrary{patterns}
\usetikzlibrary{shapes,arrows,positioning}
\usetikzlibrary{angles,quotes}
\usetikzlibrary{calc}
\usetikzlibrary{fit}
\usetikzlibrary{plotmarks}
\usetikzlibrary{decorations}
\tikzset{every picture/.style={>=stealth}}
\tikzset{three sided left/.style={
        draw=none,
        xshift=\pgflinewidth,
        append after command={
            [shorten <= -0.5\pgflinewidth]
            ([shift={(-1.5\pgflinewidth,-0.5\pgflinewidth)}]\tikzlastnode.north east) edge ([shift={( 0.5\pgflinewidth,-0.5\pgflinewidth)}]\tikzlastnode.north west) 
            ([shift={( 0.5\pgflinewidth,-0.5\pgflinewidth)}]\tikzlastnode.north west) edge ([shift={( 0.5\pgflinewidth,+0.5\pgflinewidth)}]\tikzlastnode.south west)            
            ([shift={( 0.5\pgflinewidth,+0.5\pgflinewidth)}]\tikzlastnode.south west) edge ([shift={(-1.0\pgflinewidth,+0.5\pgflinewidth)}]\tikzlastnode.south east)
        }}}
        
\tikzset{three sided right/.style={
        draw=none,
        xshift=-\pgflinewidth,
        append after command={
            [shorten <= -0.5\pgflinewidth]
            ([shift={( 1.5\pgflinewidth,-0.5\pgflinewidth)}]\tikzlastnode.north west) edge ([shift={(-0.5\pgflinewidth,-0.5\pgflinewidth)}]\tikzlastnode.north east) 
            ([shift={(-0.5\pgflinewidth,-0.5\pgflinewidth)}]\tikzlastnode.north east) edge ([shift={(-0.5\pgflinewidth,+0.5\pgflinewidth)}]\tikzlastnode.south east)            
            ([shift={(-0.5\pgflinewidth,+0.5\pgflinewidth)}]\tikzlastnode.south east) edge ([shift={( 1.0\pgflinewidth,+0.5\pgflinewidth)}]\tikzlastnode.south west)
        }}}

\usepackage{pgfplots}
\usepgfplotslibrary{polar}
\pgfplotsset{
  compat=newest, 
  width=0.60\columnwidth,    
  height=0.42\columnwidth,   
  plot coordinates/math parser=false,
  standard/.style={
    axis equal,
    axis line style=help lines,
    axis x line=center,
    axis y line=center,
    axis z line=center},
    grid style={dashed,gray},
    minor grid style={dotted,gray},
    major grid style={dotted,gray},
    ylabel absolute, ylabel style={yshift=-0.1cm},
    xlabel absolute, xlabel style={yshift=0.25cm}
}

\makeatletter
 \pgfdeclarepatternformonly[\tikz@pattern@color,\LineSpace,\LineWidth]{my horizontal lines}%
    {\pgfpointorigin}{\pgfqpoint{100pt}{1pt}}{\pgfqpoint{100pt}{\LineSpace}}%
    {
        \pgfsetcolor{\tikz@pattern@color}
        \pgfsetlinewidth{\LineWidth}
        \pgfpathmoveto{\pgfqpoint{0pt}{0.5pt}}
        \pgfpathlineto{\pgfqpoint{100pt}{0.5pt}}
        \pgfusepath{stroke}
    }
 
 \pgfdeclarepatternformonly[\tikz@pattern@color,\LineSpace,\LineWidth]{my vertical lines}%
    {\pgfpointorigin}{\pgfqpoint{1pt}{100pt}}{\pgfqpoint{\LineSpace}{100pt}}%
    {
        \pgfsetcolor{\tikz@pattern@color}
        \pgfsetlinewidth{\LineWidth}
        \pgfpathmoveto{\pgfqpoint{0.5pt}{0pt}}
        \pgfpathlineto{\pgfqpoint{0.5pt}{100pt}}
        \pgfusepath{stroke}
    } 
 
 \pgfdeclarepatternformonly[\tikz@pattern@color,\LineSpace,\LineWidth]{my grid}%
    {\pgfqpoint{-1pt}{-1pt}}{\pgfqpoint{\LineSpace}{\LineSpace}}
    {\pgfqpoint{\LineSpace}{\LineSpace}}%
    {
        \pgfsetcolor{\tikz@pattern@color}
        \pgfsetlinewidth{\LineWidth}
        \pgfpathmoveto{\pgfqpoint{0pt}{0pt}}
        \pgfpathlineto{\pgfqpoint{0pt}{\LineSpace + 0.1pt}}
        \pgfpathmoveto{\pgfqpoint{0pt}{0pt}}
        \pgfpathlineto{\pgfqpoint{\LineSpace + 0.1pt}{0pt}}
        \pgfusepath{stroke}
    }
 
 \pgfdeclarepatternformonly[\tikz@pattern@color,\LineSpace,\LineWidth]{my north east lines}
    {\pgfqpoint{-\LineWidth}{-\LineWidth}}{\pgfqpoint{\LineSpace}{\LineSpace}}
    {\pgfqpoint{\LineSpace}{\LineSpace}}%
    {
        \pgfsetcolor{\tikz@pattern@color}
        \pgfsetlinewidth{\LineWidth}
        \pgfpathmoveto{\pgfqpoint{-\LineWidth}{-\LineWidth}}
        \pgfpathlineto{\pgfqpoint{\LineSpace + 0.1pt}{\LineSpace + 0.1pt}}
        \pgfusepath{stroke}
    }
 
 \pgfdeclarepatternformonly[\tikz@pattern@color,\LineSpace,\LineWidth]{my north west lines}
    {\pgfqpoint{-\LineWidth}{-\LineWidth}}{\pgfqpoint{\LineSpace}{\LineSpace}}
    {\pgfqpoint{\LineSpace}{\LineSpace}}%
    {
        \pgfsetcolor{\tikz@pattern@color}
        \pgfsetlinewidth{\LineWidth}
        \pgfpathmoveto{\pgfqpoint{-\LineWidth}{\LineSpace}}
        \pgfpathlineto{\pgfqpoint{\LineSpace + 0.1pt}{-\LineWidth}}
        \pgfusepath{stroke}
    }
 
 \pgfdeclarepatternformonly[\tikz@pattern@color,\LineSpace,\LineWidth]{my crosshatch}%
    {\pgfqpoint{-1pt}{-1pt}}{\pgfqpoint{\LineSpace}{\LineSpace}}
    {\pgfqpoint{\LineSpace}{\LineSpace}}%
    {
        \pgfsetcolor{\tikz@pattern@color}
        \pgfsetlinewidth{\LineWidth}
        \pgfpathmoveto{\pgfqpoint{\LineSpace + 0.1pt}{0pt}}
        \pgfpathlineto{\pgfqpoint{0pt}{\LineSpace + 0.1pt}}
        \pgfpathmoveto{\pgfqpoint{0pt}{0pt}}
        \pgfpathlineto{\pgfqpoint{\LineSpace + 0.1pt}{\LineSpace + 0.1pt}}
        \pgfusepath{stroke}
    }
 
 \pgfdeclarepatternformonly[\tikz@pattern@color,\LineSpace,\PointSize]{my dots}%
    {\pgfqpoint{-\LineSpace*0.25}{-\LineSpace*0.25}}
    {\pgfqpoint{\LineSpace*0.25}{\LineSpace*0.25}}
    {\pgfqpoint{\LineSpace*0.75}{\LineSpace*0.75}}%
    {
        \pgfsetcolor{\tikz@pattern@color}
        \pgfpathcircle{\pgfqpoint{0pt}{0pt}}{\PointSize}
        \pgfusepath{fill}
    }
    
\def\calcLength(#1,#2)#3{%
\pgfpointdiff{\pgfpointanchor{#1}{center}}%
             {\pgfpointanchor{#2}{center}}%
\pgf@xa=\pgf@x%
\pgf@ya=\pgf@y%
\FPeval\@temp@a{\pgfmath@tonumber{\pgf@xa}}%
\FPeval\@temp@b{\pgfmath@tonumber{\pgf@ya}}%
\FPeval\@temp@sum{(\@temp@a*\@temp@a+\@temp@b*\@temp@b)}%
\FProot{\FPMathLen}{\@temp@sum}{2}%
\FPround\FPMathLen\FPMathLen5\relax
\global\expandafter\edef\csname #3\endcsname{\FPMathLen}
}    
    
\makeatother
 
\newdimen\LineSpace
\newdimen\PointSize
\newdimen\LineWidth
\tikzset{
    line space/.code={\LineSpace=#1},
    line space=3pt
}
\tikzset{
    point size/.code={\PointSize=#1},
    point size=.5pt
}
\tikzset{
    pattern line width/.code={\LineWidth=#1},
    pattern line width=.4pt
}

\tikzset{antenna/.style={insert path={-- coordinate (ant#1) ++(0,0.25) -- +(135:0.25) + (0,0) -- +(45:0.25)}}}

\pgfdeclaredecoration{discontinuity}{start}{
  \state{start}[width=0.5\pgfdecoratedinputsegmentremainingdistance-0.5\pgfdecorationsegmentlength,next state=first wave]
  {}
  \state{first wave}[width=\pgfdecorationsegmentlength, next state=second wave]
  {
    \pgfpathlineto{\pgfpointorigin}
    \pgfpathmoveto{\pgfqpoint{0pt}{\pgfdecorationsegmentamplitude}}
    \pgfpathcurveto
        {\pgfpoint{-0.25*\pgfmetadecorationsegmentlength}{0.75\pgfdecorationsegmentamplitude}}
        {\pgfpoint{-0.25*\pgfmetadecorationsegmentlength}{0.25\pgfdecorationsegmentamplitude}}
        {\pgfpoint{0pt}{0pt}}
    \pgfpathcurveto
        {\pgfpoint{0.25*\pgfmetadecorationsegmentlength}{-0.25\pgfdecorationsegmentamplitude}}
        {\pgfpoint{0.25*\pgfmetadecorationsegmentlength}{-0.75\pgfdecorationsegmentamplitude}}
        {\pgfpoint{0pt}{-\pgfdecorationsegmentamplitude}}
}
\state{second wave}[width=0pt, next state=do nothing]
  {
    \pgfpathmoveto{\pgfqpoint{0pt}{\pgfdecorationsegmentamplitude}}
    \pgfpathcurveto
        {\pgfpoint{-0.25*\pgfmetadecorationsegmentlength}{0.75\pgfdecorationsegmentamplitude}}
        {\pgfpoint{-0.25*\pgfmetadecorationsegmentlength}{0.25\pgfdecorationsegmentamplitude}}
        {\pgfpoint{0pt}{0pt}}
    \pgfpathcurveto
        {\pgfpoint{0.25*\pgfmetadecorationsegmentlength}{-0.25\pgfdecorationsegmentamplitude}}
        {\pgfpoint{0.25*\pgfmetadecorationsegmentlength}{-0.75\pgfdecorationsegmentamplitude}}
        {\pgfpoint{0pt}{-\pgfdecorationsegmentamplitude}}
    \pgfpathmoveto{\pgfpointorigin}
}
  \state{do nothing}[width=\pgfdecorationsegmentlength,next state=do nothing]{
    \pgfpathlineto{\pgfpointdecoratedinputsegmentlast}
  }
  \state{final}
  {
    \pgfpathlineto{\pgfpointdecoratedpathlast}
  }
}

\newcommand{\pgfextractangle}[3]{%
    \pgfmathanglebetweenpoints{\pgfpointanchor{#2}{center}}
                              {\pgfpointanchor{#3}{center}}
    \global\let#1\pgfmathresult  
}

%% file: abstract.tex
This paper investigates how to mitigate the impact of both co-channel interference and adjacent channel interference (ACI) on vehicle-to-vehicle (V2V) broadcast communication by scheduling and power control.  
The optimal
joint scheduling and power control problem, with the objective to
maximize the number of connected vehicles, is formulated as a mixed
integer programming problem with a linear objective and a quadratic constraint.  From the joint formulation,
we derive (a) the optimal scheduling problem for fixed transmit
powers as a Boolean linear programming problem and (b) the optimal
power control problem for a fixed schedule as a mixed integer linear
programming problem. Optimal schedules and power values can,
for smaller-size instances of the problem, be computed by solving first~(a) and
then~(b). To handle larger-size instances of the problem, we propose heuristic scheduling
and power control algorithms with reduced computational complexity. Simulation
results indicate that the heuristic scheduling algorithm yields
significant performance improvements compared to the baseline block-interleaver scheduler 
and that performance is further improved by the heuristic power
control algorithm. Moreover, the heuristic algorithms perform close
to the optimal scheme for small instances of the problem.

%% file: sec01_introduction.tex
\section{Introduction}
\label{sec:introduction}

\begin{figure}[t]	
	\centering	
	\vspace*{0.5cm}
	\input{Drawings/vueDropping.tex}
	\caption{System model}\label{drawing:SystemModel}
	\vspace*{0.7cm}
	\input{Drawings/acir.tex}
	\caption{Received power spectral density at receiving VUE $j$.}\label{drawing:ACI}
\end{figure}

\subsection{Motivation}

\reviewerB{
Recently, vehicle-to-vehicle (V2V) communication have captured great attention due to its potential to improve traffic safety, effective driving assistance and intelligent transport systems. The safety-critical information, such as cooperative awareness messages (CAMs) and decentralized environmental notification messages (DENM) \cite{Araniti1}, requires spreading safety related messages among surrounding vehicles either in a periodic or event triggered way.

Conveying safety critical messages in V2V networks have different requirements compared to conventional cellular communication systems. First, disseminating safety critical messages generally rely on broadcast protocols and often comes with a stringent requirement on reliability, which can be achieved if the signal to interference and noise ratio (SINR) exceeds a certain threshold\cite{Wanlu2016}. Secondly, low latency is an important requirement which restricts the possibilities for retransmissions. Moreover, retransmissions are cumbersome in a broadcast communication scenario. }

A key determining factor of reliability of a communication link is received interference power. There are two main types of interference: co-channel interference (CCI) and adjacent channel interference (ACI). The difference between these two lies in the frequency slot in which interferer transmits. CCI occurs when the interferer is transmitting on the same time-frequency slot as the intended transmitter. On the other hand, ACI occurs when the interferer is transmitting in the same timeslot, but on a nearby frequency slot.




ACI is mainly due to the nonlinearities in the power amplifier in the transmitter, which causes the transmitted spectrum to spread beyond what was intended. An example of ACI is illustrated in \figref{drawing:SystemModel} and  \figref{drawing:ACI}, where the receiver $j$ is decoding signals from transmitter $i$. \comments{Take the received signal from transmitter $i$ as an example.} Although transmitter $k$ is using a different frequency band, the signal to interference ratio $\mathrm{SIR}_{i,j}$ of receiver $j$ while decoding the signal from transmitter $i$ is limited by ACI from transmitter $k$. 
\reviewerB{
ACI is typically not a problem in a cellular communication network, since interference is dominated by CCI due to spectrum re-usage. Additionally, ACI is a significant problem in near-far situations only, i.e., when the interfering signal has much higher power than the desired one, see \figref{drawing:ACI}. In a cellular setting, ACI will be relative small in the uplink, if power control is used to equalize the received powers, and in the downlink, if users associate with the closest \gls{BS}.

However, it is known that V2V channel power gains are quite dynamic: measurements indicate that blocking vehicles can introduce high penetration losses\cite{Abbas2,Taimoor,Meireles1,He1}. Hence, a transmitting vehicle need to use a high transmit power to reach a vehicle that is blocked by other vehicles, and this will cause a near-far situation at vehicles that are not blocked. Moreover, unlike CCI, the received ACI is hard to cancel using interference cancellation techniques \cite{Hong1}. Therefore, ACI is a key factor in determining the performance in V2V communication.} \reviewerA{Not surprisingly, ACI-unaware schedulers might have bad performance in the presence of ACI. Indeed, we will see an example of a reasonable ACI-unaware scheduler in Sec.~\ref{sec:PerformanceEvaluation} that is quite suboptimal when VUEs are multiplexed in frequency.}

\subsection{State of the Art}
\reviewerB{
As pointed out above, ACI is typically not a problem in traditional cellular communication uplink/downlink scenarios. Therefore, vast majority of the scheduling and power control literature focuses upon reducing CCI alone \cite{v2vsch1,v2vsch2,v2vsch3}. \comments{However in a V2V communication scenario with dedicated spectrum, where the number of VUEs is less than the number frequency-timeslots, we can allocate separate frequency slots to VUEs, thereby nullifying the CCI.} However, in the absence of CCI, V2V broadcast communication performance is mainly limited by ACI \cite{AnverICC}.  In \cite{Albasry1}, the authors analyze the impact of ACI for device-to-device (D2D) communication, for various user densities and transmit powers, and conclude that ACI indeed causes outage problems when the user density is high. Similar conclusions have been made in \cite{Li1}, where the impact of ACI from cellular uplink to D2D communication is analyzed. In \cite{Angelakis1}, authors experimentally assess the throughput degradation due to ACI in an OFDM based communication system 802.11a, and conclude that ACI impact is indeed significant. Similar studies have been done upon 802.11b/g/n/ac in \cite{aciw1,aciw2,aciw3}. The impact of ACI when different communication technologies coexist in adjacent frequency bands have been extensively studied in \cite{acib1,acib2,acib3,acib4}. In \cite{Xia1}, the authors assess the performance degradation due to ACI when two LTE base stations are deployed in adjacent frequency channels. 

In V2V with carrier-sense multiple access (CSMA) medium access control (MAC), a potential transmitter may falsely assume that the channel is busy due to the ACI from a transmitter tuned to an adjacent channel, which causes the transmitter to defer its transmission resulting in delays\cite{Campolo1,Campolo2}. Additionally, in \cite{Campolo2}, the authors analyze both physical layer and MAC layer impacts of ACI in \gls{VANETs}. Our previous work \cite{AnverICC} studies the impact of ACI in V2V broadcast communication. 
}

\subsection{Contributions}

Our goal is to find scheduling and power control algorithms to maximize the number of connected vehicles in a V2V broadcast communication scenario. 
The scheduling and power control is made by a centralized unit (e.g., a \gls{BS}, roadside infrastructure node, or a special vehicle) based on slowly-varying channel state information (pathloss and shadowing), and communication between vehicles is direct (i.e, not via an uplink-downlink arrangement or via intermediate nodes).
By this, we increase the mutual awareness of the state (position, speed, heading, etc.) of the connected VUEs, which in turn improves vehicular safety. 
We make the following contributions to achieve this goal:
\begin{enumerate}
	\item The impact of ACI in V2V broadcast communication is evaluated.
	\item We formulate the joint scheduling and power control problem to maximize the number of successful links as a mixed integer quadratically constrained programming (MIQCP) problem. From the joint problem, we derive a pure scheduling problem (for fixed transmit powers) as a Boolean linear programming (BLP) problem and a pure power control problem (for a fixed schedule) as a mixed integer linear programming (MILP) problem. To the best of our knowledge, we are the first to formulate ACI-aware scheduling and power control problems. For small instances of the joint problem, we compute a numerically optimal solution for scheduling by solving the BLP problem and then compute a numerically optimal power values by solving the MILP problem.
	\item Due to the NP-hardness of the above scheduling problem, we suggest a block interleaver scheduler (BIS), which requires only the position indices of the VUEs.
	\item We also propose a heuristic scheduling algorithm with polynomial time complexity. The simulation results show the promising performance of the heuristic algorithm, compared to the BIS and optimal scheduler.
	\item Due to the NP-hardness of the optimal power control problem, we propose a heuristic power control algorithm as an extension of our previous work in \cite{AnverICC}. The simulation results show that the proposed algorithm further improves the performance compared to equal power.  
\end{enumerate}

\subsection{Notation and Outline} 

We use the following notation throughout the paper. Sets are denoted by calligraphic letters, e.g., $\mathcal{X}$, with $|\mathcal{X}|$ denoting its cardinality, and $\emptyset$ indicate an empty set. Lowercase and uppercase letters, e.g., $x$ and $X$, represent scalars. Lowercase boldface letters, e.g., $\xab$, represent a vector where $\xa_{i}$ is the $i$th element and $|\xab|$ is its dimensionality. The uppercase boldface letters, e.g., $\mathbf{X}$, denote matrices where $X_{i,j}$ indicates the $(i,j)^\mathrm{th}$ element. \comments{The subscript $[\cdot]_{i,j}$ also represents $(i,j)^\mathrm{th}$ element of a matrix. The superscript $(\cdot)^{\mathsf{T}}$ stands for the transposition, and $\mathbf{1}$ and $\zeroM$ represent the all-ones column vector and the zero column vector, respectively. Unless otherwise specified, vector and matrix inequalities are interpreted element-wise. We use$\!\!\mod\!(a,b)$  for the remainder of $a$ when divided by $b$.} The notations $\lceil \cdot \rceil$, and $\lfloor \cdot \rfloor$, $\round{\cdot}$ represents ceil, floor, and round operations, respectively.

\reviewerB{The rest of the paper is organized as follows. We discuss system model and ACIR model in Section~\ref{sec:preliminaries}. Section~\ref{sec:system_model} formulates optimal scheduling and power control as an optimization problem. Sections~\ref{sec:SchedulingAlgorithms} and 
\ref{sec:heuristic:power:control} describes scheduling algorithms and power control algorithms, respectively, with lower computational complexity than the optimum joint approach. Finally, we discuss numerical results in Section~\ref{sec:PerformanceEvaluation}, draw conclusions in Section~\ref{sec:Conclusions}, and describe future work in Section~\ref{sec:future works}.}


%% file: Drawings/vueDropping.tex
\begin{tikzpicture}[rec/.style={rounded corners,inner sep=10pt,draw}]
\node(VUEi) at (0,0) [rec]{VUE $i$} ;
\node(VUEj) at (4,0) [rec]{VUE $j$} ;
\node(VUEk) at (7,0) [rec]{VUE $k$} ;
\draw[bend left,->,color=green] (VUEi) to node[above,black]{Desired link} (VUEj);
\draw[bend right,->,color=red,thick] (VUEk) to node[above,black]{Interference link} (VUEj);
\draw[bend left,->,color=green] (VUEi) to node[below,black]{$H_{i,j}$} (VUEj);
\end{tikzpicture}

%% file: Drawings/acir.tex
\pgfdeclarelayer{bg}    
\pgfsetlayers{bg,main}
\tikzstyle{gArea} = [pattern=north west lines, draw=green,pattern color=green]
\tikzstyle{rArea} = [pattern=north east lines, draw=red,pattern color=red]
\tikzstyle{lStyle} = [dash dot,thick,black!50]
\begin{tikzpicture}
\begin{scope}[yshift=-4cm,scale = 1.2]{Receiver}
\draw[thick,->] (-.25,0) -- (7,0) node[anchor=north east]{Frequency};
\draw[thick,->] (0,-.25) -- (0,3.5);
\node[anchor=west,rotate=90] at (-.2cm,0.1) {Rx Power (dBm/Hz)};
\begin{pgfonlayer}{bg}
\draw[gArea] (.2,0) {[rounded corners] -- ++(30:1.5) -- ++(0.1,1) -- node(A){} ++(1,0) --++(0.1,-1) -- ++(-30:1.5)} -- cycle;
\draw[rArea,xshift=4cm] (-3.3,0)  {[rounded corners] -- ++(30:3.8) -- ++(0.2,1) -- node(B){} ++(1,0) --++(0.2,-1)}  {[opacity=0]  -- ++(-30:1) -- ++(0,-1.38)} -- cycle;
\end{pgfonlayer}
\draw[lStyle,dash phase=40pt] (B) --  ++(-3.6cm,0) node(Bl){};
\draw[lStyle] ($(A)+(4.5,-.8)$) node(Adr){}  -- ++(-5.5cm,0) node(Al){};
\draw[<->,thick,black!50] ($(Bl)+(0.2,-1pt)$) -- node[left,black]{ACIR} ++(0,-1.9);

\draw[lStyle,dash phase=13pt] ($(A)$)  -- ++(4.5cm,0) node(Ar){};
\draw[<->,thick,black!50] ($(Ar)+(-.1,-.05)$) -- node[right,black]{$\mathrm{SIR}_{i,j}$} ++(0,-.7);

\node[above,text width=2cm, align=center]  at ($(A)+(0,1.2)$) {Received signal from transmitter $i$};
\node[above,text width=2cm, align=center]  at ($(B)+(0,0)$) {Received signal from transmitter $k$};
\end{scope}
\end{tikzpicture}

%% file: sec02_system_model.tex

\section{Preliminaries}
\label{sec:preliminaries}


\begin{table}[t] 
	\caption{Key Mathematical Symbols}\label{table_notation}
	\renewcommand{\arraystretch}{1.3}
	\begin{tabular}[t]{cl}
		\hline		Symbol  &   Definition  \\
		\hline 
		$N$ & Number of VUEs \\
		$F$ & Number of frequency slots \\
		$T$ & Number of timeslots \\
		$\mathcal{T}_j$ & Set of intended transmitters for VUE $j$  \\
		$\mathcal{R}_i$ & Set of intended receivers for VUE $i$  \\
		$P_{i,t}$ & Transmit power of VUE $i$ on an RB in timeslot $t$ \\
		$\Pmax$ & Maximum transmit power of a VUE \\
		$H_{i,j}$ & Average channel power gain from VUE $i$  to VUE $j$ \\
		$\acirf$ & ACI from frequency slot $f'$ to frequency slot $f$ \\ 
		$X_{i,f,t}$ &  Indicate if VUE $i$ is scheduled to transmit in RB $(f,t)$ \\
		$Y_{j,f,t}$ & Indicate if VUE $j$ receives packet successfully in RB $(f,t)$ \\
		$\Upsilon_{i,j,t}$ & \parbox[t]{7cm}{SINR of the packet from VUE  $i$ to VUE $j$ in timeslot $t$}\\
		$\Gamma_{j,f,t}$ & SINR of the packet received by VUE  $j$ in RB $(f,t)$\\
		$\gammaT$ & \parbox[t]{7cm}{SINR threshold to declare a link as successful}\\
		$\sigma^2$ & Noise variance in an RB \\
		\hline
	\end{tabular}
\end{table}

\subsection{System Model}  \comments{TODO: Explain more about slow varying CSI}
\reviewerB{The key mathematical symbols are summarized in Table \ref{table_notation}. We consider a network of $N$ VUEs, where the set of VUEs is denoted by $\Ns \triangleq \{1,2,\ldots,N\}$. We indicate a transmitting VUE as VUE $i$, receiving VUE as VUE $j$, and interfering VUE as VUE $k$ as illustrated in \figref{drawing:SystemModel}. The average channel power gain from VUE $i$ to VUE $j$, which takes into account pathloss and large-scale fading, is denoted $H_{i,j}$. We assume, without loss of generality, that
VUE $i$ wants to transmit its packet to all VUEs in the set $\mathcal{R}_i \subset \Ns$, and VUE $j$ wants to receive packets from all VUEs in the set $\mathcal{T}_j = \{i: j \in \mathcal{R}_i\}$. We note that unicast communication is the special case when $|\mathcal{R}_i| \leq 1 \quad \forall\,i \in \Ns$. 

The total bandwidth for transmission is divided into $F$ frequency slots and the total time duration into $T$ timeslots. A time-frequency slot is also called a resource block (RB) \cite[section 6.2.3]{36.211}. We assume that a VUE can transmit its packet using a single RB. Each VUE wants to broadcast a safety message (to the VUEs in the corresponding set $\mathcal{R}$) within $T$ timeslots. Hence, the latency constraint and time-slot duration determines $T$. Given a reliability constraint and the statistics of the small-scale fading, we can compute a SINR threshold $\gammaT$ such that packets are guaranteed to be received with the required error probability if the average received SINR is equal or greater than $\gammaT$ \cite[Lemma 1]{Wanlu2016}. 

We assume that a centralized controller schedules and power control all VUEs. A base station (BS) or a VUE can act as the centralized controller. We also assume that the average channel power gain (i.e., pathloss and large-scale fading) between the VUEs are known to the centralized controller. The small-scale fading can vary on a very short time scale, on the order of milliseconds, while changes in pathloss and large-scale fading are typically small for 100~ms, even at highway speeds. It is therefore more reasonable to assume knowledge of average channel power gains (slow channel state information) than instantaneous channel gains (fast channel state information). The pathloss and large-scale fading is measured by the individual VUEs and reported to the centralized controller.}

\subsection{ACI Model}
\label{subsec:acir_model}

\begin{figure}[t]	
	\centering	
	\includegraphics{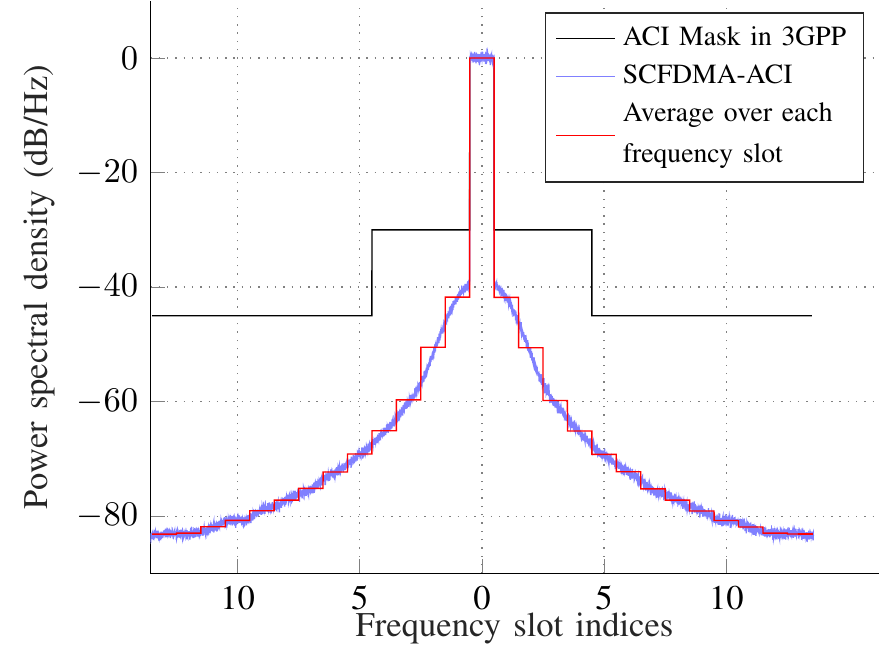}
	\caption{Inverse ACIR model} 
\label{drawing:acir_model}
\end{figure}

The ACI caused by a transmitter depends mainly upon the power amplifier, the coding and modulation scheme, and clipping threshold \cite{tommyVTC}. In \cite{Joao1}, the authors propose a two-stage low pass FIR filter method to reduce ACI in V2V communication. However, in order to find out a standard ACI model for single carrier frequency division multiple access (SCFDMA) signal, we did extensive simulations and the result for 1\% clipping threshold is shown as blue colored curve in \figref{drawing:acir_model}. The red-colored step curve in the same figure shows the SCFDMA ACI averaged over each frequency slot. The black step curve in \figref{drawing:acir_model} is the ACI mask specified for uplink by 3GPP \cite{36.942}, which is incidentally quite similar to the IEEE~802.11p mask~\cite{802.11p}.

A parameter named adjacent channel interference ratio (ACIR) is widely used to measure the ACI \cite[section 17.9]{DahlmanParkvallSkold}. As illustrated in \figref{drawing:ACI}, ACIR is defined as the ratio between the average in-band received power from interferer $k$ to the average received out of band power from interferer $k$'s signal in the frequency band allocated for transmitter $i$.
Let $\mathbf{\acir} \in \mathbb{R}^{F \times F}$ be the element-wise inverse ACIR matrix, i.e., $\acirf$ is the ratio between the received power on the frequency slot $f$ and the received power on the frequency slot $f'$, when a transmitter sends a packet on frequency slot $f'$. Observe that $\mathbf{\acir}$ is a Toeplitz matrix. The mask specified by 3GPP \cite{36.942} is as follows,
\begin{align}
\acirf &= \left\{ \begin{array}{lll}
1, & \mbox{$f'=f $} \\
10^{-3}, & \mbox{$1 \leq |f'-f| \leq 4 $} \\
 10^{-4.5}, & \mbox{otherwise} \\
\end{array} \right.
\end{align}

The scenario $f'=f$ in the above equation implies that VUEs are allocated within the same RB, in which case the interference would be CCI instead of ACI. 


%% file: sec02b_problemFormulation_ES.tex
\section{Joint Scheduling and Power Control} \label{sec:system_model}


\subsection{Constraint Formulation}
\label{sec:constraint:formulation}


In this section we will make the constraint on transmit power and scheduling mathematically precise. The objective is to maximize the number of successful links, which is done indirectly by introducing SINR constraints on as many possible desired links, i.e., the links $\{(i, j): i\in\mathcal{N}, j\in \mathcal{R}_i\}$.

\subsubsection{Transmit power constraint}
We define the matrix $\mathbf{P} \in \mathbb{R}^{N \times T }$ where $P_{i,t}$ is the transmit power of VUE $i$, if scheduled in timeslot $t$. The value of $P_{i,t}$ is constrained by the maximum transmit power of a VUE $\Pmax$ in an RB, i.e.,
\begin{equation}
0 \leq P_{i,t} \leq  \Pmax \qquad \forall\,i,t  \label{constrPower}
\end{equation}



\subsubsection{SINR constraint}
Let us define $\mathbf{\Gamma} \in \{0,1\}^{N \times F \times T}$ with $\Gamma_{j,f,t}$ as the received SINR of VUE $j$ in RB $(f,t)$, which can be computed as
\begin{equation}
 \Gamma_{j,f,t} = \frac{ S_{j,f,t}} {\sigma^2 +   I_{j,f,t} },
\end{equation}
where $S_{j,f,t}$ is the desired signal power, $\sigma^2$ is the noise variance, and $I_{j,f,t}$ is the interference power. We will show how to compute the signal and interference powers in Section~\ref{sec:S:I:computation} below. We note that focusing on the SINR of a certain receiving VUE~$j$ in an RB~$(f,t)$ allows us to state the joint scheduling and power control problem as an MIQCP problem, whereas a formulation using the SINR for specific transmitter-receiver pair would result in an harder problem as shown in Appendix~\ref{Appendix:powerControlFormulation}.  

The SINR constraint for a successful link, i.e., $\Gamma_{j,f,t} \geq \gammaT$, can be rewritten as
$S_{j,f,t} \geq \gammaT(\sigma^2 +   I_{j,f,t})$, or equivalently
\begin{equation}
S_{j,f,t}(1+\gammaT) \geq \gammaT(\sigma^2 +   I_{j,f,t} + S_{j,f,t})
\end{equation}
which in turn is equivalent to
\begin{equation}
S_{j,f,t} - \gammaTbar (I_{j,f,t} + S_{j,f,t}) \ge \gammaTbar \sigma^2,   \label{constraintSingleLink1}
\end{equation}
where $\gammaTbar \triangleq\gammaT/(1+\gammaT)$. However, it might not be possible to fulfill this condition for all receivers $j$ in all RBs $(f, t)$. To select which combinations of $j$, $f$, and $t$ to enforce this condition, we use the matrix $\mathbf{Y}\in\{0,1\}^{N\times F\times T}$, where
\begin{equation}
\label{Ydefinition}
Y_{j,f,t} \triangleq
\begin{cases}
1, & \text{\eqref{constraintSingleLink1} is enforced}\\
0, & \text{otherwise}
\end{cases}
\end{equation}

We can combine~\eqref{constraintSingleLink1} and~\eqref{Ydefinition} into a single constraint as
\begin{equation}
\label{constraintMatrix2}
S_{j,f,t} - \gammaTbar (I_{j,f,t} + S_{j,f,t})  \ge \gammaTbar
\sigma^2 - \eta(1-Y_{j,f,t})\quad\forall j, f, t    
\end{equation}
where $\eta$ is a sufficiently large number to make~\eqref{constraintMatrix2} hold whenever $Y_{j,f,t}=0$, regardless of the schedule and power allocation. It is not hard to show that $\eta = \gammaTbar(NP^{\mathrm{max}}+\sigma^2)$ is sufficient. 

\subsubsection{Scheduling constraints}
Let $\mathbf{X} \in \{0,1\}^{N \times F \times T }$ be the scheduling matrix defined as
\begin{equation}
X_{i,f,t} \triangleq 
\begin{cases}
1, & \text{VUE $i$ is scheduled in RB $(f,t)$ } \\
0, & \text{otherwise} \\
\end{cases}
\end{equation}
We limit a VUE scheduling to at most one RB in a timeslot, since scheduling in multiple RBs in a timeslot reduces available transmit power in an RB, and spreads interference across multiple RBs. Hence we add the following constraint,
\begin{align}
	\sum_{f = 1}^{F} X_{i,f,t} \leq 1  \quad \forall i,\, t   \label{constraintnRBsForAVUE}.
\end{align}

We recall that VUE $j$ is interested in decoding packets from the VUEs in the set $\mathcal{T}_j$. If we set $Y_{j,f,t}=1$, we want the SINR for receiver VUE $j$ in RB $(f,t)$ to be above $\gammaT$ for a transmitter VUE in $\mathcal{T}_j$. It then makes sense to not to allow more than one VUE in $\mathcal{T}_j$ to transmit in RB $(f,t)$, which is enforced by the following constraint,
\begin{equation}
	\sum_{i \in \mathcal{T}_j} X_{i,f,t} \leq 1 + N(1-Y_{j,f,t}) \quad \forall j,\, f,\, t.  \label{constraintnVUEsInAnRB}
\end{equation}
Note that the above constraint is always satisfied when $Y_{j,f,t}=0$, since $|\mathcal{T}_j|\le N$. However, when $Y_{j,f,t}=1$ then~\eqref{constraintnVUEsInAnRB} implies that at most one VUE in $\mathcal{T}_j$ can transmit in RB $(f,t)$ and CCI can therefore only be due to VUEs in the set $\mathcal{N}\setminus\mathcal{T}_j$, a fact that will be used in \eqref{definitionI} below.

\subsubsection{Computation of $S_{j, f, t}$ and $I_{j,f,t}$}
\label{sec:S:I:computation}
It follows from the scheduling constraints~\eqref{constraintnRBsForAVUE} and~\eqref{constraintnVUEsInAnRB} that the desired signal power $S_{j, f, t}$ and interference power $I_{j,f,t}$ needed in the SINR constraint~\eqref{constraintMatrix2} can be computed as
\begin{align}
  S_{j,f,t}   &   =\sum_{i \in \mathcal{T}_j}   X_{i,f,t} P_{i,t} H_{i,j}  \label{definitionS}~,   \\
I_{j,f,t} &= \sum_{k \in \Ns \setminus \mathcal{T}_j}  X_{k,f,t}  P_{k,t} H_{k,j}  \nonumber \\
    &\hspace*{0.5cm} +   \overset{F}{\underset{\substack{f' = 1 \\ f' \neq f}}{\sum}}  \sum_{k \in\mathcal{N}} \acirf X_{k,f',t}  P_{k,t} H_{k,j} ~,  \label{definitionI}   
\end{align}
We note that the first term in \eqref{definitionI} is CCI from VUEs not in $\mathcal{T}_j$ and that the second term is ACI from all transmitting VUEs. 

\subsection{Problem Formulation}  \label{subsec:problemFormulation}
We define a link as a transmitter-receiver pair $(i,j)$, and we say that the link $(i,j)$ is successful if at least one transmission from VUE~$i$ to VUE~$j$ is successful during the scheduling interval, i.e., that the SINR condition (\ref{constraintSingleLink1}) is satisfied for at least one RB $(f,t)$ where $f\in\{1, 2, \ldots, F\}$ and $t\in\{1, 2, \ldots, T\}$. We introduce the matrix $\mathbf{Z}\in\{0,1\}^{N\times N}$, where, for all $i, j$,
\begin{align}
Z_{i,j} &\triangleq \min\{1,   \sum_{t=1}^T\sum_{f=1}^F X_{i,f,t}Y_{j,f,t}\}  \label{ZijDefinition} \\
&=
\begin{cases}
1, & \text{link $(i,j)$ is successful}     \\
0, & \text{otherwise}
\end{cases}   
\end{align}
where the minimum in~\eqref{ZijDefinition} is required to not to count successful links between VUE~$i$ and VUE~$j$ more than once.

The overall goal is to maximize the number of connected VUE pairs, i.e., to maximize the objective function
\begin{equation}
\label{jointProblemStatement}
J(\mathbf{X}, \mathbf{Y}, \mathbf{P})\triangleq \sum_{i=1}^N\sum_{\substack{j=1\\j\neq i}}^N Z_{i,j}
\end{equation}
subject to the constraints (\ref{constraintnVUEsInAnRB}), (\ref{constraintnRBsForAVUE}), (\ref{constrPower}), (\ref{constraintMatrix2}), and (\ref{ZijDefinition}). However, since $J$ is nonlinear in the binary matrices $\mathbf{X}$ and $\mathbf{Y}$, direct optimization of $J$ is cumbersome. We will therefore formulate an equivalent optimization problem which is simpler to solve. To this end, let us define two auxiliary matrices $\mathbf{V}\in\mathbb{R}^{N\times N\times F\times T}$ and $\mathbf{W}\in\mathbb{R}^{N\times N}$, where, for all $i, j$,
\begin{align}
\label{definitionV}
V_{i,j,f,t} &\in\{v\in\mathbb{R}: v \le X_{i,f,t}, v\le Y_{j,f,t}\},\\
W_{i,j} &\in\{w\in\mathbb{R}: w \le 1, w\le \sum_{t=1}^T\sum_{f=1}^F V_{i,j,f,t}\}.  \label{definitionW}
\end{align}
Now, for any fixed $\mathbf{X}$, $\mathbf{Y}$, it follows from~\eqref{definitionV} that 
\begin{equation}
V_{i,j,f,t}^\star = \max V_{i,j,f,t} = \min\{X_{i,f,t},Y_{j,f,t}\} = X_{i,f,t}Y_{j,f,t}.  \label{defVstar}
\end{equation}
The last equality in the above equation follows from the fact that both $X_{i,f,t}$ and $Y_{j,f,t}$ are Boolean.  Moreover, it follows from~\eqref{definitionW} and~\eqref{ZijDefinition} that if $V_{i,j,f,t} = V_{i,j,f,t}^\star$, then
\begin{equation}
 \max W_{i,j} = \min\{1,\sum_{t=1}^T\sum_{f=1}^F V^{\star}_{i,j,f,t}\} = Z_{i,j}.  
\end{equation}

Hence, for any fixed $\mathbf{X}$, $\mathbf{Y}$, $\mathbf{P}$ we can compute $J(\mathbf{X}, \mathbf{Y}, \mathbf{P})$ as the optimal value of objective of 
\begin{maxi!}
	{\mathbf{V}, \mathbf{W}}{\sum_{i=1}^N \sum_{\substack{j=1\\ j\neq i}}^N W_{i,j}}
	{\label{eq:7}}{J(\mathbf{X}, \mathbf{Y}, \mathbf{P}) =}
	\addConstraint{\eqref{definitionV}, \eqref{definitionW}} \nonumber
\end{maxi!}

Putting everything together, we arrive at the optimization problem   
	\begin{subequations} \label{blp}
	\begin{align}
	&\   \max\limits_{\mathbf{P},\Xb,\mathbf{Y},\mathbf{V},\mathbf{W}}  \overset{N}{\underset{i=1}{\sum}}  \sum_{j \in \mathcal{R}_i}  W_{i,j}  \label{blp_obj} \\
	&\mbox{s.t. } \nonumber \\
	& \underset{i \in \mathcal{T}_j}{\sum}   X_{i,f,t} P_{i,t} H_{i,j}   - \gammaTbar  \overset{F}{\underset{f' = 1}{\sum}}  \overset{N}{\underset{k=1}{\sum}}  \acirf X_{k,f',t} P_{k,t} H_{k,j} \hspace*{1.5cm}    \nonumber \\
	&   \hspace*{0.2cm}\geq \gammaTbar \sigma^2 - \gammaTbar(N\Pmax+\sigma^2) (1-Y_{j,f,t}) \hspace*{0.3cm} \forall \, j,f,t    \label{blpConstr1} \\[5pt]    
	&W_{i,j}   \leq  \overset{T}{\underset{t=1}{\sum}} \overset{F}{\underset{f=1}{\sum}}    V_{i,j,f,t}  \hspace*{3.2cm}  \forall\, i,j   \label{blp_constr_Wij} \\
	&W_{i,j} \leq 1 \hspace*{5cm}  \forall\, i,j  \\
	&V_{i,j,f,t} \leq X_{i,f,t}   \hspace*{4.1cm}   \forall\, i,j,f,t  \\
	&V_{i,j,f,t} \leq Y_{j,f,t}    \hspace*{4.2cm}  \forall\, i,j,f,t  \\
	&   \sum_{i \in \mathcal{T}_j} X_{i,f,t} \leq 1 + N(1-Y_{j,f,t})   \hspace*{1.6cm}   \forall\, j,\, f,\, t  \\
	&\sum_{f = 1}^{F} X_{i,f,t} \leq 1  \hspace*{4.3cm}   \forall\, i,\, t \\
	&\  0 \leq P_{i,t} \leq \Pmax   \hspace*{4cm}  \forall\, i,t  \\				
	&\mathbf{X},\mathbf{Y} \in \{0,1\}^{N \times F \times T }	\label{blp_booleanConstr}  \hspace*{2.8cm} 	 \\
	&\Pb\in\mathbb{R}^{N\times T} \\
	&\mathbf{V}\in\mathbb{R}^{N\times N\times F\times T} \\
	&\mathbf{W}\in\mathbb{R}^{N\times N}
	\end{align}
\end{subequations}

\reviewerB{
Here are some of the key observations regarding the above problem formulation:
\begin{enumerate}[label=(\roman*)]

\item We see that the problem~\eqref{blp} has linear objective and linear constraints except the constraint (\ref{blpConstr1}), which is quadratic. We call such a problem a MIQCP problem. Moreover, the problem~\eqref{blp} is noncovex even after relaxing the Boolean constrains for $\Xb$ and $\Yb$ as proved in Appendix~\ref{Appendix-nonConvexity}. 
Since there are $2NFT$ Boolean variables and $(NT+N^2FT+N^2)$ continuous variables in our power control problem formulation, we see that the worst-case computational complexity is $\mathcal{O}(\frac{(NT+N^2FT+N^2)^3 2^{2NFT}}{\log (NT+N^2FT+N^2)})$. The complexity $2^{2NFT}$ is due to fixing $2NFT$ Boolean variables, and the complexity $\frac{(NT+N^2FT+N^2)^3}{\log (NT+N^2FT+N^2)}$ is for solving each of the resulting \gls{LP} problem using an interior point method \cite{Florian1}. 

\item \label{problem:SchedulingOptimal} The problem formulation (\ref{blp}) can be translated into a  scheduling alone problem by fixing all power values $P_{i,t}$. The resulting problem is a BLP problem, with worst case computational complexity $\mathcal{O}(\frac{(N^2FT+N^2)^3 2^{2NFT}}{\log (N^2FT+N^2)})$ since there are $2NFT$ Boolean variables and $(N^2FT+N^2)$ continuous variables.

\item \label{problem:PowercontrolOptimal} 
The problem formulation~\eqref{blp} can be translated into a power control alone problem for an arbitrary scheduling matrix $\Xb$. That is, we fix the scheduling matrix $\Xb$ and optimize over $\Pb$ with the following modified objective function,
\begin{equation} \label{nearOptimalPC}
 \max\limits_{\mathbf{P},\mathbf{Y},\mathbf{V},\mathbf{W}}   \sum_{i=1}^{N}  \sum_{j \in \mathcal{R}_i}  W_{i,j}  -\beta  \sum_{t=1}^{T} \sum_{i=1}^{N} P_{i,t},
\end{equation}
where $\beta$ is the weight of the total power consumption in the objective, in order to achieve our secondary goal of minimizing the total power consumption. We note that if $\beta \leq 1/(N T \Pmax)$, then the sum power minimization will not affect our primary objective of maximizing the total number of successful links. Furthermore, we can change constraint \eqref{blp_constr_Wij} to $W_{i,j} \leq \sum\limits_{(f,t):X_{i,f,t}=1} Y_{j,f,t}$, thereby avoiding the need for variable $\mathbf{V}$.

Observe that the problem of finding the optimal power values is NP-hard as proved in \cite[Lemma 1]{AnverICC}. The worst-case computational complexity is $\mathcal{O}(\frac{(N^2 + NT)^3 2^{NFT}}{\log (N^2 + NT)})$, where the complexity $2^{NFT}$ is due to fixing $NFT$ Boolean variables, and the complexity $\frac{(N^2 + NT)^3}{\log (N^2 + NT)}$ is for solving each of the resulting \gls{LP} problem using an interior point method \cite{Florian1}.

\item The problem formulation (\ref{blp}) allows for full-duplex communication, i.e., a VUE can simultaneously transmit and receive. Half-duplex communication can be enforced by adding the following constraint,\footnote{High values of self-interference channel gain (i.e., diagonal values of matrix $\Hb$), will effectively force the solution to be half-duplex. However, this could cause numerical issues for the solver. Therefore, if half-duplex communication is desired, using constraint~\eqref{constrHalfDuplexCondition} and setting the self-interference channel power gain values to zero, (i.e., $H_{i,i}=0 ~~\forall\,i$) is highly recommended due to numerical issues.}

\begin{equation} \label{constrHalfDuplexCondition}
Y_{i,j,f,t} \leq (1-X_{j,f',t}) \quad \forall\,i,\,j,\,f,\,f',\,t
\end{equation}

\item The optimization problem in (\ref{blp}) can be reformulated to maximize the minimum number of successful links for a VUE, instead of the total number of successful links. By doing this, at least $L^*$ links are guaranteed to be successful for any VUE. This is done by changing the objective function in (\ref{blp_obj}) to
\begin{equation}
L^* = \max\limits_{\mathlarger{\mathbf{P},\Xb,\mathbf{Y},\mathbf{V},\mathbf{W},L}}   L
\end{equation} 
and adding an extra constraint,
\begin{equation}
\sum_{\substack{j=1 \\ j \neq i}}^{N}  Z_{i,j} \geq L  \qquad \forall\,i
\end{equation}

\item Furthermore, we note that the problem formulation in (\ref{blp}) can also be used for unicast communication by setting $\mathcal{R}_i$ to a singleton set containing the intended receiver of VUE $i$, for all $i \in \Ns$. This way, we are reducing the number of constraints in the problem and, therefore, also the computational complexity.

\end{enumerate}

}

%% file: sec03_contributions.tex
\section{Scheduling Algorithms} \label{sec:SchedulingAlgorithms}

For the scheduling problem, without considering any power control, we set the transmit power of all VUEs to $\Pbar$, where, $0\leq \Pbar \leq \Pmax$. For the sake of scheduling all available RBs, we define VUE $0$ as a dummy VUE with zero transmit power. Hence, scheduling VUE $0$ to an RB indicate that no VUE is scheduled in that RB. 

Let us define the matrix $\mathbf{U} \in \{0,1,\dots,N\}^{F \times T}$ to represent scheduled VUEs in an $F \times T$ RBs matrix. That is, $U_{f,t}$ is the VUE index scheduled in RB $(f,t)$. Fundamentally, scheduling is the process of allocating VUEs in available RBs, which is equivalent to populating the $\Ub$ matrix with appropriate VUE indices, as illustrated in \figref{schedulingMatrix}. Once we have computed $\Ub$, the matrix $\Xb$ can be computed as follows,
\begin{equation}
  X_{i,f,t} =
  \begin{cases}
    1,  &  U_{f,t} = i \\
    0,  &  \text{otherwise}
  \end{cases}.
  \label{eqUtoX}
\end{equation}

%

\subsection{Block Interleaver Scheduler (BIS)}

%


The algorithm is summarized in Algorithm \ref{Alg:BIS}. The approach here is to insert each VUE index exactly once in $\mathbf{U}$. Clearly, this is impossible if $N> FT$, i.e., when  there are more VUEs than available RBs. For the time being, we will assume that $N\le FT$ and treat the $N> FT$ case later in this Section. Moreover, we will assume that $N>T$, since the scheduling problem is trivial otherwise; we can simply schedule the VUEs in separate timeslots, which removes all ACI and CCI interferences.  

If $N > T$, then we need to multiplex VUEs in frequency, which results in ACI. To reduce the ACI problem, we strive to use as few frequency slots as possible and to space the frequency slots as far apart as possible. Since $T$ VUEs can be scheduled per frequency slot, the smallest required number of frequency slots is $\tilde{F} = \lceil N/T\rceil$. Clearly, $\tilde{F}\le F$, since we assume that $N\le FT$. The selected frequency slots are put in the vector $\fab\in\{1, 2, \ldots, F\}^{\tilde{F}}$. For BIS, we will use the frequency slots
\begin{equation}
\label{computefab}
\fa_{k} = 1 + \left\lceil (k-1) \frac{F-1 }{\tilde{F}-1}\right\rfloor, \qquad k = 1, 2, \ldots, \tilde{F}.
\end{equation}
We note that
$\fa_1 = 1 < \fa_2 <\cdots< \fa_{\tilde{F}}=F$, and  it can be shown that~\eqref{computefab} maximizes the minimum distance between any two consecutive frequency slots, i.e., maximizes
\begin{equation}
\label{eq:2}
\min_{l\in\{1, 2, \ldots, \tilde{F}-1\}} |\fa_{l+1}-\fa_{l}|.
\end{equation}

We initialize $\Ub=\zeroM^{F \times T}$. Then, given $\fab$, BIS starts by filling the rows of $\mathbf{U}$ in the natural way, i.e., row $\fa_1$ with VUE indices $1, 2,\ldots, T$, row $\fa_2$ with indices $T+1, T+2, \ldots, 2T$, and so on. To (possibly) improve the scheduler, the nonzero rows of $\mathbf{U}$ are then permuted with a block interleaver $\Pi$; which is equivalent to permuting $\fab$ with the block interleaver $\Pi$ before filling in the rows of $\mathbf{U}$.

Now we explain the block interleaver $\Pi$ used to permute $\fab$. Our block interleaver is same as the one specified in 3GPP \cite[section 5.1.4.2.1]{36.211}. We define $\fab' = \Pi(\fab,w)$ as the output $\fab'$ of a block interleaver with width $w \in \mathbb{N}$ and input vector $\fab$. The block interleaver writes $\fab$ row-wise in a matrix with width $w$, padding with zeros if necessary, then reads $\fab'$ from the matrix column-wise ignoring zeros. Observe that if $w=1$, then the block interleaver output is same as the input, i.e., $\fab' = \fab$. The width of the block interleaver $w$ is an input to this algorithm. 
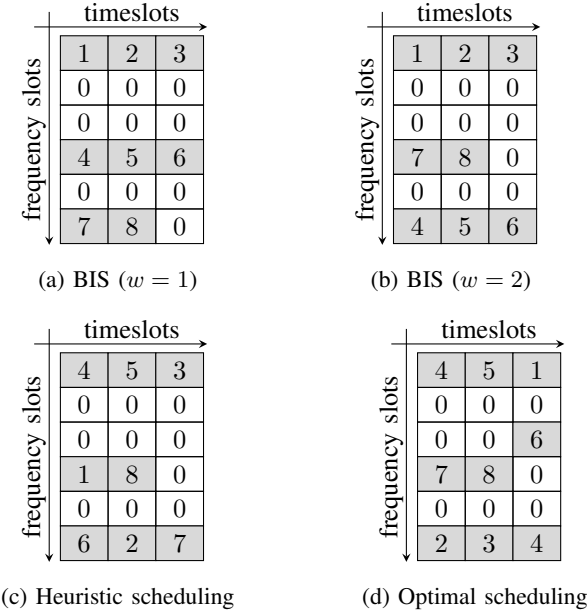
\begin{figure}[t]	  	
	
	
	\begin{subfigure}[t]{0.45\linewidth}
		\centering
		\input{Drawings/naturalSchedulingMatrix.tex}
		\caption{BIS ($w=1$)} \label{naturalSchedulingMatrix}
	\end{subfigure}	\hfill
	\begin{subfigure}[t]{0.45\linewidth}
		\centering
		\input{Drawings/OBI_schedulingMatrix.tex}
		\caption{BIS ($w=2$)}  \label{OBI_schedulingMatrix}
	\end{subfigure} \hfill
	\medskip
	
	\begin{subfigure}[t]{0.45\linewidth}
		\centering
		\input{Drawings/heuristic_schedulingMatrix.tex}
		\caption{Heuristic scheduling}  \label{heuristic_schedulingMatrix}
	\end{subfigure}	\hfill
	\begin{subfigure}[t]{0.45\linewidth}
		\centering
		\input{Drawings/optimal_schedulingMatrix.tex}
		\caption{Optimal scheduling} \label{optimal_schedulingMatrix}
	\end{subfigure} \hfill
	
	\caption{Example of scheduling 8 VUEs in $6 \times 3$ RBs. VUEs are placed on a convoy with inter vehicular distance 48.6\,m  } \label{schedulingMatrix}	  
\end{figure}

\begin{algorithm}[H]
	\caption{Block Interleaver Scheduler (BIS)}
	\label{Alg:BIS}
	\begin{algorithmic}[1]
		\Require{$\{N, F, T, w\}$}
		\Ensure{$\mathbf{X}$}
    	\State $\tilde N = \min\{\lfloor NT/2\rfloor, N, FT\}$		
		\State $\tilde F = \lceil \tilde{N}/T\rceil$
		\State Compute $\fab$ and $\nab$ from~\eqref{computefab} and~\eqref{computenab} 
		\State $\fab'= \Pi(\fab, w)$
		\State $\Ub=\zeroM^{F \times T}$
		\State $k = 1$
		\For{$l =1:|\fab'|$}
		\State $f' = \fa_{l}'$
		\For{$t =1:T$}
		\If{$k\le |\nab|$}
		\State $U_{f',t}=\na_k$
		\State $k = k+1$
		\EndIf
		\EndFor
		\EndFor
		\State Compute $\mathbf{X}$ from $\mathbf{U}$ using (\ref{eqUtoX})
	\end{algorithmic}
\end{algorithm}

As an example, when $N=8,F=6, T=3, w=1$, we compute $\fab' = \fab = [1,4,6]$, and schedule VUEs accordingly as shown in \figref{schedulingMatrix}\,(a). Similarly, \figref{schedulingMatrix}\,(b) shows the result when $w=2$ and the computed $\fab' = [1,6,4]$. We present the results for various values of $w$ in Section \ref{subsec:SimulationResults}. 

Now let us treat the case when $N > FT$. One way to handle this case is to schedule only $\tilde{N}\le FT$ of the $N$ VUEs. For BIS, we put the selected VUEs in the vector $\mathbf{n}\in\{1, 2, \ldots, N\}^{\tilde{N}}$, where
\begin{equation}
n_{k} = 1 + \left\lceil (k-1) \frac{N-1 }{\tilde{N}-1}\right\rfloor, \qquad k
= 1, 2, \ldots, \tilde{N}.  \label{computenab}
\end{equation}
We note that if $\tilde{N}=N$, then $\mathbf{n} =[1,2,\ldots, N]$. Hence, the two cases $N \le FT$ and $N > FT$ can be unified by letting $\tilde{N} = \min\{N, FT\}$ and $\tilde{F} = \lceil \tilde{N}/T\rceil$. 

However, if $T=1$, then it is never advantageous to schedule more than $\lfloor N/2\rfloor$ VUEs in the half-duplex case. To understand why, we note that since we have $\tilde{N}$ transmitters and $N-\tilde{N}$ receivers, the maximum number of successful links we can ever hope for is $\tilde{N}(N-\tilde{N}) = (N/2)^2 - (\tilde{N} - N/2)^2$, which is maximized when $\tilde{N}=\min\{\lfloor N/2\rfloor, F\}$. Scheduling more than $\lfloor N/2\rfloor$ VUEs will not increase the number of possible links (due to half-duplex criteria), but increase ACI. The final, unifying, calculation of $\tilde{N}$ in Algorithm~\ref{Alg:BIS} is therefore $\tilde{N} = \min\{\lfloor TN/2\rfloor, N, FT\}$ and $\tilde{F} = \lceil \tilde{N}/T\rceil$, which covers all cases of $N$, $F$, and $T$.

%


\subsection{Heuristic Scheduling Algorithm}

The approach taken here is to loop through all RBs and schedule either a real or dummy VUE to each RB. The scheduling decision is taken in a greedy fashion. That is, we strive to schedule the best possible VUE to the RB under the assumption that the schedule for all previous RBs is fixed. The resulting schedule can schedule a VUE, zero, one, or multiple times, as opposed to BIS, which schedules all real VUEs exactly once (if there are enough RBs, $FT \geq N$ and $T>1 $).

The heuristic algorithm is executed in two steps: 1) Determine the RB scheduling order, 2) Use this order to sequentially visit the RBs and schedule VUEs.

Now we explain the first step, i.e., the procedure to compute the scheduling order $\fab$ for frequency slots. We note that $\fab$ is a permutation of $\{1,2,\dots,F\}$, which can be chosen in $F!$ possible ways. We compute $\fab$ using a greedy algorithm as shown in Algorithm \ref{heuristic1Algorithm3.1}. That is, while constructing $\fab$, our priority is to spread out the consecutive scheduling frequency slots in order to minimize the received ACI. Therefore, in each iteration, we are scheduling a frequency slot with minimum received ACI from all the scheduled frequency slots. Therefore, we always start scheduling from the first frequency slot, i.e., $\fa_1 = 1$, then we find out the next frequency slot $\fa_2$ as the unscheduled frequency slot with minimum received ACI from $\fa_1$. We repeat this process \comments{, i.e., finding the unscheduled frequency slot with minimum received ACI from all the scheduled frequency slots, }until all frequency slots are chosen. Finding the frequency slot with minimum received ACI from all the scheduled frequency slots is actually impossible, since we do not know yet which VUE is going to be scheduled in the RBs and its transmit power. Therefore, we compute the ACI in an unscheduled frequency slot by assuming unit transmit power and unit channel gain from all interferers. If there are multiple unscheduled frequency slots with the same minimum affected ACI, then the frequency slot having maximum average distance from all the scheduled frequency slots is chosen. If there is still a tie, then the $\max$ value is chosen as shown in Algorithm \ref{heuristic1Algorithm3.1}, line 5. This way, $\fa_2=F$ is ensured for a typical ACIR model.


Next we explain the second step, i.e., finding out the VUE to schedule in an RB. The algorithm is shown in Algorithm \ref{heuristic1Algorithm}. Given an RB to schedule, first we compute the total number of successful links upon scheduling each VUE in the chosen RB, we then pick the VUE which would maximize this quantity. Observe that VUE $0$ (the dummy VUE) can be scheduled to an RB, which, of course, means that no real VUE is scheduled. Counting the number of \textit{for} loops and the operations on lines 11 and 12 in Algorithm \ref{heuristic1Algorithm}, we see that the heuristic scheduling is a polynomial time algorithm with the worst case computational complexity $\mathcal{O}(NFT(FT+N^2))$.

The result of the scheduling when $N=8,F=6, T=3$, is shown in \figref{schedulingMatrix}\,(c), when VUEs are placed on a one lane road, with equal distances \davg \,(refer to Table \ref{table:Simulation_Parameters}) to the neighboring VUEs, and by assuming zero shadow loss. Note that in this example VUE $4$ is scheduled twice.

\begin{algorithm}[!t]
	\renewcommand{\thealgorithm}{2.1}
	\caption{Computation of scheduling order $\fab$}
	\label{heuristic1Algorithm3.1}
	\begin{algorithmic}[1]
		\Require{$\{F,\mathbf{\acir}\}$}
		\Ensure{$\fab$}
		\State $\fa_1 = 1  $
		\State $\Fs = \{2,3,\dots,F\}  $
		\For{$l=2:F$}
		\State $\Gs = \argmin\limits_{f \in \Fs} \sum\limits_{l'=1}^{l-1} A_{\fa_{l'},f}$
		\State $\fa_{l} = \max \left\{ \argmax\limits_{f \in \Gs} \sum\limits_{l'=1}^{l-1} |f-f_{l'}| \right\}$
		\State $\Fs = \Fs \setminus \fa_{l}$
		\EndFor		
	\end{algorithmic}
\end{algorithm}

\begin{algorithm}[!t]
	\renewcommand{\thealgorithm}{2.2}
	\caption{Heuristic Scheduling Algorithm}
	\label{heuristic1Algorithm}
	\begin{algorithmic}[1]
		\Require{$\{N,F,T,\Hb,\mathbf{\acir},\Pb,\gammaT,\sigma^2 \}$}
		\Ensure{$\Xb$}
		\State $\mathbf{X} = \zeroM^{N \times F \times T}, \quad \Ub = \zeroM^{F \times T} $ 
		\State Compute $\fab$ using Algorithm \ref{heuristic1Algorithm3.1}  \\
		// Schedule RBs in the order specified by $\fab$
		\For{$l = 1:F $}
		\State  $f = \fa_{l}$
		\For{$t = 1:T $}
		\State // Schedule VUE in RB $(f,t)$
		\For{$i=0:N$}
		\State $U_{f,t} = i$
		\State Compute $\Xb$ from $\Ub$ using (\ref{eqUtoX})
		\State Compute $\Zb$ for $\mathbf{X}$ using (\ref{ZijDefinition})		
		\State  {\raggedright  $s_i = \overset{N}{\underset{m=1}{\sum}}   \sum\limits_{j \in \mathcal{R}_m} Z_{m,j}$} 
		\EndFor
		\State $U_{f,t}={\underset{i}{\argmax}} \{s_i\} $ 
		\EndFor
		\EndFor
		\State Compute $\Xb$ from $\Ub$ using (\ref{eqUtoX})				
	\end{algorithmic}
\end{algorithm}




%% file: Drawings/naturalSchedulingMatrix.tex

	\tikzset{
		table/.style={
			matrix of nodes,
			row sep=-\pgflinewidth,
			column sep=-\pgflinewidth,
			nodes={
				rectangle,
				draw=black,
				align=center,
				text width = 0.4cm,    
				minimum height=0.3cm,
				anchor=south,
			},
			nodes in empty cells,
			row 1 column 1/.style={
				nodes={fill=gray!30}
			},
			row 1 column 2/.style={
			nodes={fill=gray!30}
			},
			row 1 column 3/.style={
			nodes={fill=gray!30}
			},
			row 4 column 1/.style={
			nodes={fill=gray!30}
			},
			row 4 column 2/.style={
				nodes={fill=gray!30}
			},
			row 4 column 3/.style={
				nodes={fill=gray!30}
			},
			row 6 column 1/.style={
				nodes={fill=gray!30}
			},
			row 6 column 2/.style={
				nodes={fill=gray!30}
			}
					}
	}
	\begin{tikzpicture}
	
	\matrix (mat) [table]
	{
		$1$ & $2$ & $3$ \\
		$0$ & $0$ & $0$ \\
		$0$ & $0$ & $0$ \\
		$4$ & $5$ & $6$ \\
		$0$ & $0$ & $0$ \\
		$7$ & $8$ & $0$ \\
%
%
	};
\def\a{0.9cm}
\draw[ ->] (-1.3,1.5) -- (1,1.5);
\node (RB) at (0,1.7cm) {timeslots};

\def\a{1.1cm}
\def\av{1.4cm}
\draw[ ->] ($(-\a,\av)+(0,0.3cm)$) -- (-\a,-\av);
\node[rotate=90] (RB) at (-1.3cm, 0) {frequency slots};

	\end{tikzpicture}

%% file: Drawings/OBI_schedulingMatrix.tex

	\tikzset{
		table/.style={
			matrix of nodes,
			row sep=-\pgflinewidth,
			column sep=-\pgflinewidth,
			nodes={
				rectangle,
				draw=black,
				align=center,
				text width = 0.4cm,    
				minimum height=0.3cm,
				anchor=south,
			},
			nodes in empty cells,
			row 1 column 1/.style={
				nodes={fill=gray!30}
			},
			row 1 column 2/.style={
			nodes={fill=gray!30}
			},
			row 1 column 3/.style={
			nodes={fill=gray!30}
			},
			row 4 column 1/.style={
			nodes={fill=gray!30}
			},
			row 4 column 2/.style={
				nodes={fill=gray!30}
			},
			row 6 column 3/.style={
				nodes={fill=gray!30}
			},
			row 6 column 1/.style={
				nodes={fill=gray!30}
			},
			row 6 column 2/.style={
				nodes={fill=gray!30}
			}
					}
	}
	\begin{tikzpicture}
	
	\matrix (mat) [table]
	{
		$1$ & $2$ & $3$ \\
		$0$ & $0$ & $0$ \\
		$0$ & $0$ & $0$ \\
		$7$ & $8$ & $0$ \\
		$0$ & $0$ & $0$ \\
		$4$ & $5$ & $6$ \\
%
%
	};
\def\a{0.9cm}
\draw[ ->] (-1.3,1.5) -- (1,1.5);
\node (RB) at (0,1.7cm) {timeslots};

\def\a{1.1cm}
\def\av{1.4cm}
\draw[ ->] ($(-\a,\av)+(0,0.3cm)$) -- (-\a,-\av);
\node[rotate=90] (RB) at (-1.3cm, 0) {frequency slots};

	\end{tikzpicture}

%% file: Drawings/heuristic_schedulingMatrix.tex

	\tikzset{
		table/.style={
			matrix of nodes,
			row sep=-\pgflinewidth,
			column sep=-\pgflinewidth,
			nodes={
				rectangle,
				draw=black,
				align=center,
				text width = 0.4cm,    
				minimum height=0.3cm,
				anchor=south,
			},
			nodes in empty cells,
			row 1 column 1/.style={
				nodes={fill=gray!30}
			},
			row 1 column 2/.style={
			nodes={fill=gray!30}
			},
			row 1 column 3/.style={
			nodes={fill=gray!30}
			},
			row 4 column 1/.style={
			nodes={fill=gray!30}
			},
			row 4 column 2/.style={
				nodes={fill=gray!30}
			},
			row 6 column 1/.style={
				nodes={fill=gray!30}
			},
			row 6 column 2/.style={
				nodes={fill=gray!30}
			},
			row 6 column 3/.style={
			nodes={fill=gray!30}
			}
		}
	}
	\begin{tikzpicture}
	
	\matrix (mat) [table]
	{
		$4$ & $5$ & $3$ \\
		$0$ & $0$ & $0$ \\
		$0$ & $0$ & $0$ \\
		$1$ & $8$ & $0$ \\
		$0$ & $0$ & $0$ \\
		$6$ & $2$ & $7$ \\
%
%
	};
\def\a{0.9cm}
\draw[ ->] (-1.3,1.5) -- (1,1.5);
\node (RB) at (0,1.7cm) {timeslots};

\def\a{1.1cm}
\def\av{1.4cm}
\draw[ ->] ($(-\a,\av)+(0,0.3cm)$) -- (-\a,-\av);
\node[rotate=90] (RB) at (-1.3cm, 0) {frequency slots};

	\end{tikzpicture}

%% file: Drawings/optimal_schedulingMatrix.tex

	\tikzset{
		table/.style={
			matrix of nodes,
			row sep=-\pgflinewidth,
			column sep=-\pgflinewidth,
			nodes={
				rectangle,
				draw=black,
				align=center,
				text width = 0.4cm,    
				minimum height=0.3cm,
				anchor=south,
			},
			nodes in empty cells,
			row 1 column 1/.style={
				nodes={fill=gray!30}
			},
			row 1 column 2/.style={
			nodes={fill=gray!30}
			},
			row 1 column 3/.style={
			nodes={fill=gray!30}
			},
			row 3 column 3/.style={
			nodes={fill=gray!30}
			},
			row 4 column 1/.style={
			nodes={fill=gray!30}
			},
			row 4 column 2/.style={
				nodes={fill=gray!30}
			},
			row 6 column 1/.style={
				nodes={fill=gray!30}
			},
			row 6 column 2/.style={
				nodes={fill=gray!30}
			},
			row 6 column 3/.style={
			nodes={fill=gray!30}
			}
		}
	}
	\begin{tikzpicture}
	
	\matrix (mat) [table]
	{
		$4$ & $5$ & $1$ \\
		$0$ & $0$ & $0$ \\
		$0$ & $0$ & $6$ \\
		$7$ & $8$ & $0$ \\
		$0$ & $0$ & $0$ \\
		$2$ & $3$ & $4$ \\
%
%
	};
\def\a{0.9cm}
\draw[ ->] (-1.3,1.5) -- (1,1.5);
\node (RB) at (0,1.7cm) {timeslots};

\def\a{1.1cm}
\def\av{1.4cm}
\draw[ ->] ($(-\a,\av)+(0,0.3cm)$) -- (-\a,-\av);
\node[rotate=90] (RB) at (-1.3cm, 0) {frequency slots};

	\end{tikzpicture}

%% file: sec04_powerControl.tex
\section{Heuristic Power Control}
\label{sec:heuristic:power:control}
Since the exponentially increasing worst-case complexity of optimal power control is problematic in practice for large networks, we propose a heuristic power control algorithm which has polynomial time computational complexity. The proposed heuristic power control algorithm is an extension of our previous work on power control \cite{AnverICC} and the work of Kang Wang et al. \cite{KangWang}. All those previous works assumes $T=1$, whereas our proposed algorithm finds a power control solution for any value of $T$. The algorithm is described in Algorithm \ref{heuristicPowerControlAlgorithm}. 

The SINR $\Upsilon_{i,j,t}$ of a link $(i,j)$ during the timeslot $t$ is computed as follows,
\begin{equation}
\Upsilon_{i,j,t} = \frac{ \overset{F}{\underset{f=1}{\sum}}   X_{i,f,t} P_{i,t} H_{i,j}} {\sigma^2 +  \sum\limits_{f=1}^{F} \overset{F}{\underset{\substack{f' = 1}}{\sum}}  \overset{N}{\underset{\substack{k=1 \\ k \neq i}}{\sum}}  X_{i,f,t} \acirf X_{k,f',t} P_{k,t} H_{k,j} } \label{Gammaijt_computation1}.
\end{equation}
 The derivation of the above equation is explained in Appendix~\ref{Appendix:powerControlFormulation}. A link $(i,j)$ is successful if and only if its SINR is greater than or equals to $\gammaT$ on any timeslot, i.e., $\Upsilon_{i,j,t} \geq \gammaT$ for any $t \in \{1,2,\dots,T\}$. Our goal is to find the optimal transmit power value for each VUE in each timeslot in order to maximize the total number of successful links. The algorithm is an iterative algorithm involving two steps in each iteration. Since it may not be possible to ensure success for all links, our first step is to find the set of candidate links $\Ls$. The second step is to compute the power values $P_{i,t}$ for all VUEs in all timeslots in order to maximize the number of successful links in $\Ls$.  Therefore, we update both $\Ls$ and $P_{i,t}\,\, \forall\,i,t$ in each iteration. We terminate the algorithm, when we observe that all the links in $\Ls$ are achieving the SINR target $\gammaT$.

Now we explain the first step, i.e., the computation of $\Ls$ on each iteration.  In the first iteration, we initialize $\Ls$ to the set of all links, and in the subsequent iterations we remove some of the links from $\Ls$, thereby making $\Ls$ a nonincreasing set over iterations. We initialize all VUEs transmit power to $\Pinit$, i.e., $P_{i,t} = \Pinit\,\forall\,i,t$. We then define the variable $\Pt_{i,j,t}$ as the required transmit power of VUE $i$ during the timeslot $t$ in an iteration, so that the link $(i,j)$ would be successful in the next iteration, under the assumption that the interference remains constant\comments{ and equal to the currently measured interference}. The value of $\Pt_{i,j,t}$ is computed in each iteration as shown in Algorithm \ref{heuristicPowerControlAlgorithm}, line \ref{Alg3:line9}. If the required power for a link $(i,j)$ is more than $\Pmax$, i.e., $\Pt_{i,j,t} > \Pmax\,\,\forall\,t$, then the link $(i,j)$ is declared as a broken link. The set of broken links $\Bs$ in an iteration is computed in Algorithm \ref{heuristicPowerControlAlgorithm}, line \ref{Alg3:line10}. We find out repeatedly broken links over many iterations and remove them from the set $\Ls$ (line \ref{Alg4:line16}). 

In order to find the repeatedly broken links, a counter $C_{i,j}$ is set to count the number of iterations at which the link $(i,j)$ gets broken. We remove the link $(i,j)$ from $\Ls$ once $C_{i,j}$ reaches above a threshold $\Ct$, i.e, $C_{i,j} > \Ct$.  We observe that, the algorithm shows improved performance as we increase $\Ct$. However, higher values of $\Ct$ increases computational complexity due to more number of iterations. Moreover, the initial transmit power $\Pinit$ plays a crucial role in this algorithm. A higher value of $\Pinit$ leads to more number of broken links in the first iteration itself, meanwhile lower values lead to a slow convergence of the algorithm. By simulations, we observe that $\Pinit = \Pmax / 10$ is a reasonable value for $\Pinit$.

\begin{algorithm}[!t]
	\renewcommand{\thealgorithm}{3}
	\caption{Heuristic Power Control}
	\label{heuristicPowerControlAlgorithm}
	\begin{algorithmic}[1]
	    \Require{$\{N,F,T,\Pinit,\Pmax,\Xb,\Hb,\mathbf{\acir},\gammaT,\sigma^2 \}$}
		\Ensure{$\Pb$}
		 \State $P_{i,t} = \Pinit \quad \forall\, i, t  $  \vspace*{0.1cm}
		\State $\mathbf{C} = \zeroM^{N \times N}$ 
		\vspace*{0.3cm}
		 \State  \parbox{8.5cm}{ // set of candidate links  \\ $ \Ls = \{(i,j): \overset{T}{\underset{t=1}{\sum}} \overset{F}{\underset{f=1}{\sum}}  X_{i,f,t} >0, j \in \mathcal{R}_i \}$}   \vspace*{0.2cm}
		 \State \parbox{8.5cm}{// scheduled time-slots for VUE $i$ \\ $ \timeslots_i = \{ t: \sum\limits_{f=1}^F  X_{i,f,t} >0 \} \quad \forall\, i $}  \vspace*{0cm}
		   \State Compute SINR $\Upsilon_{i,j,t} \quad\forall\, i,j,t $ using (\ref{Gammaijt_computation1})\vspace{0.1cm} 		 
		\While{$ \exists\, (i,j) \in \Ls \,\,\mathrm{s.t.}\,\,  \Upsilon_{i,j,t} < \gammaT \quad \forall\, t$}  \label{alg4:termination condition}
		   \State // Compute the required power and broken links $\Bs$
		   \State $\tilde{P}_{i,j,t} = \frac{\gammaT}{\Upsilon_{i,j,t}} P_{i,t} \quad \forall\, (i,j) \in \Ls, t \in \timeslots_i $  \label{Alg3:line9}
		   \State $ \Bs = \{(i,j): \tilde{P}_{i,j,t} > \Pmax  \quad \forall\, t \in \timeslots_i \} $ \label{Alg3:line10}
		   \State // Increment $C_{i,j}$ and update $\Ls$
		   \State $C_{i,j} = C_{i,j} + 1 \quad \forall\, (i,j) \in \Bs$
		   \State $ \Ls = \Ls \setminus \{(i,j): C_{i,j} > \Ct \}$ \label{Alg4:line16}
		   \State $\Rsbar_i = \{j:(i,j) \in \Ls \setminus \Bs\} \quad \forall i  $   \label{Alg3:line14}
		   \State // Compute power values
		   \State $P_{i,t} = 0 \,\,\,\, \forall\, i,t $
		   \For{$i = 1:N$}  \label{Alg4:line16}
		      \While{$\Rsbar_i \neq \emptyset $}  \label{Alg4:line17}
		      	   \State $\Ks_t = \{\tilde{P}_{i,j,t} : \tilde{P}_{i,j,t}\leq \Pmax, \,\, j \in \Rsbar_i \}  \,\, \forall\, t \in \timeslots_i$ \label{Alg3:line18}
				   \State $t^\star = {\underset{t \in \timeslots_i}{\argmax}}  \left|  \Ks_t \right| $
				   \State $P_{i,t^\star} = \max  \Ks_{t^\star}$  \label{Alg3:line20}
				   \vspace{0.1cm}
				   \State $\Rsbar^{\star}_i = \{j: P_{i,t^\star}  \geq  \tilde{P}_{i,j,t^\star}   \} $
				   \State $\Rsbar_i  = \Rsbar_i \setminus \Rsbar^{\star}_i $
			   \EndWhile
		   \EndFor \vspace{0.2cm}
		   \State \parbox{8cm}{Compute SINR $\Upsilon_{i,j,t} \,\, \forall\,i,j,t $ using (\ref{Gammaijt_computation1}) with updated power values} \vspace{-0.1cm} 		   
		\EndWhile
	\end{algorithmic}
\end{algorithm}

Next we explain the second step, i.e., the computation of power values $P_{i,t}\,\, \forall\,t$, in each iteration. We compute the power values of each VUE independently. In the following, we therefore explain the power value computation of an arbitrary VUE $i$ for all timeslots $t \in \{1,2,\dots,T\}$. Let us define the set $\Rsbar_i$ as the set of intended receivers in $\Ls \setminus \Bs$ for the transmitting VUE $i$, as computed in Algorithm \ref{heuristicPowerControlAlgorithm}, line \ref{Alg3:line14}. Our goal is to make the received SINR of all the links from VUE $i$ to VUEs in $\Rsbar_i$ equal to or greater than $\gammaT$ in the next iteration, i.e., $\Upsilon_{i,j,t} \geq \gammaT ~ \forall\,j\in\Rsbar_i$ . Therefore, we compute $P_{i,t}\,\,\forall\,t$, such that the SINR values of all the links in $\Ls \setminus \Bs$ are greater or equal to $\gammaT$ on at least one of the timeslots in the next iteration, under the assumption that the interference remains constant. 

Furthermore, in order to minimize the interference to other links, we would consider allocating power to a VUE in as few number of timeslots as possible. Therefore, the power allocation to VUE $i$ involves two steps. The first step is to decide the optimal timeslot $t^\star$ to allocate power, and the second step is to compute the power value for the chosen timeslot $t^\star$. We compute $t^\star$ as the timeslot at which VUE $i$ can serve the maximum number of intended receivers in $\Rsbar_i$. For this purpose, we first compute $\Ks_t$ as the set of transmit powers for VUE $i$ that are required to serve the receivers in $\Rsbar_i$  and do not exceed $\Pmax$, as shown in Algorithm \ref{heuristicPowerControlAlgorithm}, line \ref{Alg3:line18}. Clearly, the cardinality of this set, i.e., $|\Ks_t|$, is the number of receivers that can be served during timeslot $t$ in the next iteration. Therefore, $t^{\star}$ is computed as the timeslot $t$ that maximizes $|\Ks_t|$ (i.e., $t^{\star} = \argmax_t |\Ks_t|$ ), and ties are broken arbitrarily. We compute the power value $P_{i,t^\star}$ as the maximum value in $\Ks_{t^\star}$ (which is less than $\Pmax$), as shown in Algorithm \ref{heuristicPowerControlAlgorithm}, line \ref{Alg3:line20}. Then we compute the set of receivers $\Rsbar^{\star}_i$ which are served by the allocated power $P_{i,t^\star}$, and remove those from $\Rsbar_i$, thereby making the set $\Rsbar_i$ as the set of VUEs not yet served. We repeat these two steps until the allocated transmit power $P_{i,t}$ is greater or equal to the required transmit power $\Pt_{i,j,t}$ on at least one of the timeslot $t$, for all receivers in $\Rsbar_i$.

The algorithm is convergent since maximum number of iterations possible in line \ref{alg4:termination condition} is $\Ct\left|\Ls\right|$ as proved in Lemma \ref{lemma2} in Appendix \ref{Appendix:ProvingConvergence}. Counting the number of iterations in lines \ref{alg4:termination condition}, \ref{Alg4:line16}, \ref{Alg4:line17} and computation of $\Upsilon_{i,j,t}$ in algorithm \ref{alg:power}, we see that the heuristic power control is a polynomial time algorithm with worst case computational complexity $\mathcal{O}(\Ct N^6 T)$.

%


%% file: sec05_SimulationResults.tex

\section{Performance Evaluation}    \label{sec:PerformanceEvaluation}
\begin{table}[t]
	\centering
	\caption{System Simulation Parameters}
	\label{table:Simulation_Parameters}
	\begin{tabular}{|l|l|}
		\hline
		Parameter & Value \\
		\hline
		ACIR model & 3GPP mask \\
		$\gammaT$ & $5$ dB \\
		$\Pmax$ & $24$ dBm \\
		$\Pinit$ & $\Pmax/10$  \\
		$\text{PL}_0$ & $63.3$ dB \\
		$n$ & $1.77$  \\
		$d_0$ & 10 m \\
		$\sigma_1$ & $3.1$ dB \\
		Penetration Loss & $10$ dB per obstructing VUE \\
		$\sigma^2$ & $-95.2$ dBm\\
		\davg & 48.6 m \\
		$d_{\text{min}}$ & 10 m \\
		$\beta$ & $1/(N \Pmax)$ \\
		$\eta$ & $\gammaT(N\Pmax+\sigma^2) $ \\
		$ \Ct $ & $100$ \\
		\hline
	\end{tabular}
\end{table}

\subsection{Scenario and Parameters}  \label{scenarioAndParameters}

\comments{The VUEs are numbered as per location index in the convoy as illustrated in \figref{drawing:SystemModel}.}
For the simulation purpose, we consider a platooning scenario, where $N$ VUEs are distributed on a convoy, as used in the realtime vehicular channel measurements done in \cite{Karedal}.
The distance between any two adjacent VUEs, $d$, follows a shifted exponential distribution, with the minimum distance $d_{\text{min}}$ and the average distance $d_{\text{avg}}$ \cite{Koufos1,Dewen1,Cowan1,Luttinen1}. That is, the probability density function of $d$ is given as,
\begin{equation}
  f(d) =
  \begin{cases}
    \frac{1}{d_{\text{avg}}-d_{\text{min}}}  \exp({-\frac{d-d_{\text{min}}}{d_{\text{avg}}-d_{\text{min}}}}) ,
    & d \ge d_{\text{min}} \\
    0, & \text{otherwise}
  \end{cases}
\end{equation}

Following the recommendation by 3GPP \cite[section A.1.2]{36.885} for freeway scenario, $d_{\text{avg}}$ is set to $48.6$\,m, which corresponds to 2.5 seconds for a vehicular speed of $70\,$km/h. \reviewerA{We note that the mobility is less of a concern for the time scale of the problem under study. Typically, the latency requirement is less than 100~ms, over which time the slow channel state information (i.e., pathloss and shadowing) typically does not vary significantly, even in a highway speed. Fast channel variations (i.e., small-scale fading) is accounted for in the calculation of $\gammaT$. That is, $\gammaT$ is computed from the small-scale fading statistics (not its realizations) and the reliability constraint, see \cite[Lemma 1]{Wanlu2016} for details. In other words, there is no need for an explicit mobility model to assess performance of the scheduling and power control algorithms in this paper.}

\reviewerA{We adopt the channel model \comments{and channel parameters}from \cite{Karedal}, which is a model based on the real-time measurements of V2V links at carrier frequency 5.2 GHz in a highway scenario. We note that the measurements in \cite{Karedal} are consistent with the measurements done in \cite{Abbas1,Lin1,Kunisch1}.} The pathloss in dB for a distance $d$ is computed as,
\begin{equation}
\text{PL}(d) = \text{PL}_0 + 10 n\hspace*{0.05cm}\text{log}_{10}(d/d_0) + X_{\sigma_1}
\end{equation}
where $n$ is the pathloss exponent, $\text{PL}_0$ is the pathloss at a reference distance $d_0$, and $X_{\sigma_1} $ represents the shadowing effect modeled as a zero-mean Gaussian random variable with standard deviation $\sigma_1$. The values of the channel parameters are taken from \cite{Karedal} (shown in Table~\ref{table:Simulation_Parameters}), which is based upon real-time measurements in a highway scenario. \reviewerA{The penetration loss caused by multiple obstructing vehicles is not fully understood yet. However, the penetration loss caused by a single vehicle has been widely studied. Measurements show that an obstructing truck causes 12--13\,dB \cite{Taimoor}, a bus 15--20\,dB \cite{He1}, a van 20\,dB \cite{Meireles1}, and a car 10\,dB \cite{Abbas2} penetration loss.} 
\reviewerA{To summarize, there is no widely accepted, measurement-based model for the penetration loss of multiple vehicles available in the literature. For simulations purpose, we therefore simply assume that each blocking vehicle introduce an additional attenuation of 10\,dB.} The noise variance is $-95.2\,$dBm and $\Pmax$ is $24\,$dBm as per 3GPP recommendations \cite{36.942}. We assume that $d_{\text{min}} = 10\,$m and that $\gammaT = 5\,$dB  is sufficient for a transmission to be declared as successful (i.e., that the error probability averaged over the small-scale fading is sufficiently small). Additionally, we fix $\Ct=100$, which is found to be a reasonable value for the heuristic power control algorithm. 
For the simulation purpose, the set $\mathcal{T}_j$ is chosen as the closest $\min(N-1,FT-1)$ VUEs to VUE $j$ based on the distance between the VUEs.

\reviewerB{We present results for half-duplex communication in this paper. Moreover, for simulation purposes, we use the ACI mask specified for uplink by 3GPP \cite{36.942}, since LTE uplink physical layer is a possible candidate for vehicular communication~\cite{Araniti1} upon introduction of Cellular-V2X in release~14 of the LTE standard~\cite{36.300}. Simulation results for full-duplex and the SCFDMA ACI model is available in the report \cite{AnverArchive}, but is not presented here due to space constraints.}


\subsection{Simulation Results} \label{subsec:SimulationResults}

\newcommand{\plotAextraXTickPos}{1,2,3,4,5,6,7,8,9,10,11,12,13,14,15,16,17,18,19,20}
\newcommand{\plotAxTicklabels}{2,4,6,8,10}		
\newcommand{\folderName}{2018May31}
\newcommand{\ymaxBandC}{2.05}  \newcommand{\ymaxA}{6.2}
\newcommand{\yTicksBandC}{1,1.5,2} 

\input{sec05_SimulationResults_figureA.tex}
\input{sec05_fairness_figureC.tex}

\begin{table*}[t]
	\centering
	\caption{\reviewerA{Summary of compared algorithms \\ (Performance compared when $N=20, F=20$, and $T=2$)}}
	\label{table:compared_algorithms}
	\renewcommand{\arraystretch}{1.3}
	\begin{tabular}{llllcc}
          \hline
          Line Style & Scheduling & Power control & Worst-Case Complexity & \parbox[t]{1.3cm}{\centering Performance \\ ($\bar{Z}$)} & \parbox[t]{2cm}{\centering Performance ($\bar{Z}$) \\(no ACI case)} \\[0.3cm]
          \hline
          \ref{plotsA:plot1} & BIS ($w=1$) & Equal power & $\mathcal{O}(FT+F)$ & 2.16 & 3.50\\
          \ref{plotsA:plot2} & BIS (optimized $w$) & Equal power & $\mathcal{O}(F(FT+F))$ & 2.57 & 3.50\\
          \ref{plotsA:plot2b} & \cite{Peng1} & Equal power & $\mathcal{O}(N^2FT)$ & 2.66 & 4.29\\
          \ref{plotsA:plot3}& Heuristic scheduling & Equal power & $\mathcal{O}(NFT(FT+N^2))$ & 3.36 & 3.82\\
          \ref{plotsA:plot4} & Optimal scheduling (numerical) & Equal power & $\mathcal{O}(\frac{(N^2FT+N^2)^3 2^{2NFT}}{\log (N^2FT+N^2)})$ & 3.89  & 4.79\\
\ref{plotsA:plot1} & BIS ($w=1$) & Heuristic power control & $\mathcal{O}((FT+F)\Ct N^6 T)$ & 2.63 & 3.50 \\
\ref{plotsA:plot1} & BIS ($w=1$) & Optimal power (numerical) & $\mathcal{O}((FT+F)\frac{(N^2 + NT)^3 2^{NFT}}{\log (N^2 + NT)})$ & 2.78 & 3.50 \\
  &  \multicolumn{2}{l}{Optimal joint scheduling and power control (numerical)} & $\mathcal{O}(\frac{(NT+N^2FT+N^2)^3 2^{2NFT}}{\log (NT+N^2FT+N^2)})$ & 4.58  & 4.79\\
          \hline
	\end{tabular}
\end{table*}

To measure performance, we use the following metrics
\begin{align}
&    Z_i = \sum_{j \in \mathcal{R}_i}  Z_{i,j},   \label{defZi}\\
&    \bar{Z}_i = \E[Z_i],   \label{defZibar}\\   
& \bar{Z} = \frac{1}{N}\sum_{i=1}^N  \bar{Z}_i ,  \label{defZbar}
\end{align}
where $Z_i$ is the number of successful links from VUE $i$, when VUE $i$ is transmitting a packet to all VUEs in set $\mathcal{R}_i$. The quantity $\bar{Z}_i$ is the expected value of $Z_i$, where the expectation is taken over the random quantities in the experiment, i.e., the inter-VUE distances and shadow fading. Finally, $\bar{Z}$ is the number of successful links for a VUE, averaged across all VUEs. In other words, the metric $\bar{Z}$ can be interpreted as the average number of receiving VUEs that can decode a packet from a certain VUE. Clearly, we would like to ensure that $\bar{Z}$ is sufficiently large to support the application in mind. However, to specify this minimum acceptable value of $\bar{Z}$ is out of scope of this paper.

\reviewerA{We use Gurobi solver \cite{Gurobi} for finding optimal scheduling and optimal power values, as described in \ref{problem:SchedulingOptimal} and \ref{problem:PowercontrolOptimal} in Section \ref{subsec:problemFormulation} respectively. However, we observe that Gurobi solver provides only near-optimal solution due to the numerical sensitivity of the problem, caused by high dynamic range of $\mathbf{A}$ and $\mathbf{H}$. Therefore, we refer the solutions provided by the solver as Optimal scheduling (numerical) and Optimal power (numerical), in~Figs.~\ref{resultNSuccessfulLinks}--\ref{resultTxPower}.}

Since the block interleaver width $w$ is an input parameter to BIS, we considered a class of BIS with all possible $w \in \{1,2,\dots,\Ft-1\}$. We present here the results for the optimal $w$ which maximizes $\bar{Z}$ under the assumption of equal transmit powers, shown as the blue curves marked with triangles in \figref{resultNSuccessfulLinks}. The corresponding $w$ for BIS is shown as an extra x label on top of Figs.~\ref{resultNSuccessfulLinks}(a)--(c), and we do not vary $w$ with respect to the power control algorithms.

\reviewerA{To the best of our knowledge, there is no multicast scheduling algorithm with the objective of maximizing the connectivity in the current literature. In \cite{Peng1}, authors propose a multicast scheduling algorithm to improve Quality of Services (QoS). As a benchmark, we simulate the proposed algorithm in\cite{Peng1} after modifying the objective function to maximize the connectivity (instead of improving QoS), and plotted as violet curve marked with plus in Figs.~\ref{resultNSuccessfulLinks}-\ref{resultFairness}. The proposed scheduling algorithm in \cite{Peng1} is an ACI-unaware algorithm, and performance seems to be comparable with BIS (optimized $w$). However, \cite{Peng1} assumes channel knowledge, whereas BIS does not require any channel information.}

In \figref{resultNSuccessfulLinks}, we present the result for various values of $F,\,T,\,N\,$, and various scheduling and power control algorithms. 
In Figs.~\ref{resultNSuccessfulLinks}(a)--(c), we present the results for equal power, i.e., when all VUEs transmit with the same power $\bar{P}$. We know that the performance improves as $\bar{P}$ increases, since both the signal power and the interference power are linear functions of $\bar{P}$, thereby making the SINR an increasing function of $\bar{P}$. Therefore, we set $\bar{P} = \Pmax$.
   In \figref{resultNSuccessfulLinks}(a), we plot $\bar{Z}$ by varying $T$ for a fixed $F$ and $N$. The results in \figref{resultNSuccessfulLinks}(a) clearly show that $\bar{Z}$ is severely limited by ACI when many VUEs must be multiplexed in frequency, i.e., when $T$ is small compared to $N$. This motivates the search for scheduling and power control methods to mitigate the ACI problem in this situation. We also observe that $\bar{Z}$ remains essentially constant for $T \geq 10$ due to limitations by noise power.

One way to limit the effect of ACI would be to increase $F$ (for a fixed $N$ and $T$) to allow for larger spacing of VUEs in frequency. However, the results in \figref{resultNSuccessfulLinks}(b) show that $\bar{Z}$ is only slowly increasing with $F$. On the other hand, \figref{resultNSuccessfulLinks}(b) shows that significant gains can be achieved by more advanced scheduling than using a BIS.

Moreover, for a fixed $T$ and $F$, we see in \figref{resultNSuccessfulLinks}(c) that $\bar{Z}$ is increasing with $N$, at least for the more advanced schedulers. This might be surprising at first sight; however, this effect is not unreasonable, since more receivers become available for each transmission when $N$ increases. In other words, the number of terms in the double sum in~\eqref{defZbar} increases, which tends to increase $\bar{Z}$. However, the performance flattens out for higher values of $N$ (i.e., $N \geq 20$). This is because as the network size grows, the links between VUEs that are blocked by several other VUEs become noise limited due to the penetration loss of the blocking VUEs. In this case scheduling and power control cannot improve the performance anymore.



\input{sec05_SimulationResults_figureB.tex}

As seen in Figs.~\ref{resultNSuccessfulLinks}(d)--(i), power control improves performance, but, in general, the gains are marginal for advanced schedulers. The performance gain is more significant for the BIS scheduler compared to the more advanced schedulers. This can be explained by the fact that a suboptimal schedule can be corrected to some degree by power control. Indeed, assigning zero or a very low power to a VUE effectively changes the schedule for that VUE. For instance, that the performance for BIS with $w=1$ for large $N$ is significantly improved with power control, as seen in \figref{resultNSuccessfulLinks}(f) and \figref{resultNSuccessfulLinks}(i).

It is, of course, possible to iterate between scheduling and power control. However, we have observed that this gives only marginal improvement at the price of significantly increased computational complexity. Due to space constraints, detailed results are not presented here.

In \figref{resultFairness}(a), we plot CDF of the number of successful links for a VUE, $Z_i$ defined in \eqref{defZi}, for fairness comparison between various scheduling algorithms. \reviewerA{We observe that, BIS and \cite{Peng1} perform better in terms of fairness than the heuristic scheduling algorithm, in the sense that its corresponding CDF is more steep in \figref{resultFairness}(a).} 
In \figref{resultFairness}(b), we plot the average number of successful links for each VUE, $\bar{Z}_i$ defined in \eqref{defZibar}, in a convoy of 20 VUEs. We note that VUEs in the middle of the convoy are able to successfully broadcast their packets to more number of VUEs than the VUEs on the edge of the convoy, which is logical since the VUEs in the middle have more number of close-by neighbors. Moreover, even if BIS ($w=1$) is more fair, the per-VUE performance is uniformly worse compared to the other algorithms. Except for the naturally lower $\bar{Z}_i$ for the edge VUEs, all algorithms are seen to be approximately fair.

\reviewerA{
  In Table \ref{table:compared_algorithms}, we summarize the computational complexity of the studied algorithms and the performance for a benchmark case when $N=20,~F=20$, and $T=2$. We also show the result for optimal scheduling and power control (numerical) upon solving MIQCP problem \eqref{blp}. 
  The result for scheduling algorithms (i.e., first 5 rows in Table \ref{table:compared_algorithms}) are given for the equal power control, and results for the power control algorithms (i.e., $6^\mathrm{th}$ and $7^\mathrm{th}$ rows) are given for the scheduling algorithm BIS ($w=1$). The last column in the table is the performance for no ACI case, i.e., $\mathcal{A}_{f',f}=0,~\forall\,f\neq f'$. For no ACI case, a non-overlapping scheduling (which avoids CCI) and maximum transmit power for each VUE will give the best performance. However, due to the half-duplex assumption, careful splitting the VUEs into  transmitter and receiver roles in each timeslot yields improved performance. Therefore, the improvement seen by more advanced schedulers in the last column in the table is due to this effect. Also, we note that the optimal joint scheduling and power control can more or less nullify the negative impact of ACI since its performance with and without ACI are comparable. 

It should be stressed that a scheduling and power control method that is only concerned with CCI and ignores ACI would be trivial in the case when full-duplex communication is possible and when 
$N\leq FT$: scheduling all VUEs in non-overlapping RBs and allocate maximum transmit power $\Pmax$ to all VUEs would be thought to be optimum since no CCI would occur. For half-duplex, the case is a bit more complicated. If VUE~$i$ is scheduled in timeslot $t$, then we should avoid scheduling any other VUEs in $\mathcal{R}_i$ in the same timeslot. If this is possible, the schedule is optimal (if ACI can be ignored). Indeed, all schedules in Fig.~\ref{schedulingMatrix}(a) are optimum (if ACI can be ignored) when all VUEs want to communicate with their two closest neighbors on each side. 
However, we note that ignoring ACI can lead to considerable performance loss, as the case is for the BIS ($w=1$) scheduler in  \figref{resultNSuccessfulLinks}.}


In \figref{resultTxPower}, we plot the average transmitter power values for various power control algorithms, upon fixing the scheduling algorithm as BIS with $w=1$. We observe that our proposed heuristic power control algorithm uses less transmit power compared to equal power, and close to the transmit power used by optimal power control. 

For detailed results on full-duplex and SCFDMA ACI, interested readers are directed to our report in the archive \cite{AnverArchive}. We observe that the optimal scheduling algorithm show significant performance improvement for full-duplex communication scenarios when ACIR equals to 3GPP mask. Moreover, the simulation results in the report \cite{AnverArchive} show that the order of performance for the algorithms is the same as the one presented here, regardless of the ACI model. We also plot the average transmit power values for various scheduling algorithms in \cite{AnverArchive}, and observe the similar trends. Additionally, the MATLAB code used for the simulation is shared on github \cite{AnverMatlabCode}.



%% file: sec05_SimulationResults_figureA.tex
\pgfplotstableread[col sep=comma]{\folderName/schPerformanceForEqualPowerT.csv}\datatableT
\pgfplotstableread[col sep=comma]{\folderName/schPerformanceForEqualPowerF.csv}\datatableF
\pgfplotstableread[col sep=comma]{\folderName/schPerformanceForEqualPowerN.csv}\datatableN


\pgfplotscreateplotcyclelist{colorList1}{
	{black,mark=square},
	{blue,mark=triangle},
	{violet,mark=+},
	{green,mark=star},
	{red,mark=diamond},
}

\begingroup					
\thickmuskip=0mu		

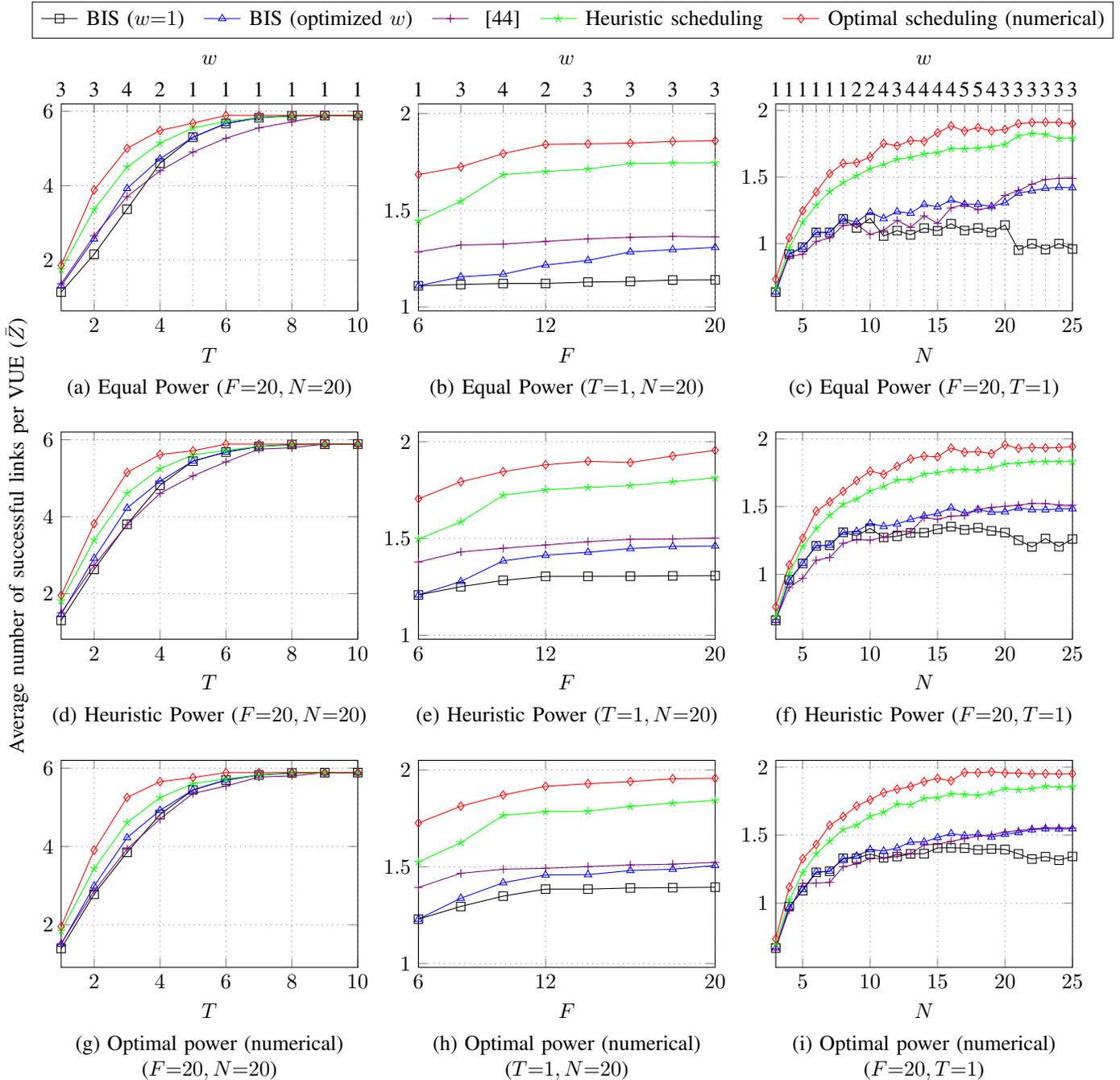
\begin{figure*}[!ht]	
	\centering	

        \begin{tikzpicture}
\begin{groupplot}[cycle list name=colorList1, xmajorgrids, ymajorgrids,
	group style={group name=my plots,group size= 3 by 3,vertical sep=2cm },
	height=5cm,width=0.3\paperwidth		
	]
	
\nextgroupplot[title=$w$, title style={yshift=2ex,},
xmin = 1,
xmax = 10,
ymax = \ymaxA,
xlabel=$T$,
xtick =  \plotAxTicklabels,
extra x ticks =  \plotAextraXTickPos,
every extra x tick/.style={
	xtick pos=right,
	xticklabel pos=right,		
	xticklabels from table={\datatableT}{W},
}
]
\addplot +[restrict expr to domain={\coordindex}{0:9}] table [x=xValues, y=sch_naturalScheduling, col sep=comma] {\folderName/schPerformanceForEqualPowerT.csv}; \label{plotsA:plot1}  
\addplot +[restrict expr to domain={\coordindex}{0:9}]  table [x=xValues, y=sch_optimizedBlockInterleaver1, col sep=comma] {\folderName/schPerformanceForEqualPowerT.csv}; \label{plotsA:plot2}
\addplot +[restrict expr to domain={\coordindex}{0:9}]  table [x=xValues, y=sch_Peng1, col sep=comma] {\folderName/schPerformanceForEqualPowerT.csv}; \label{plotsA:plot2b}
\addplot +[restrict expr to domain={\coordindex}{0:9}]  table [x=xValues, y=sch_Heuristic1, col sep=comma] {\folderName/schPerformanceForEqualPowerT.csv}; \label{plotsA:plot3}
\addplot +[restrict expr to domain={\coordindex}{0:9}]  table [x=xValues, y=sch_gurobi, col sep=comma] {\folderName/schPerformanceForEqualPowerT.csv}; \label{plotsA:plot4}

\nextgroupplot[title=$w$, title style={yshift=2ex,},
xmin = 6,
xmax = 20,
ymax = \ymaxBandC, ymin = 0.98,
ytick =  \yTicksBandC,
xlabel=$F$,
xtick =  {6,12,20},
extra x ticks =  {6,8,10,12,14,16,18,20},
every extra x tick/.style={
	xtick pos=right,
	xticklabel pos=right,		
	xticklabels from table={\datatableF}{W},
}
]
\addplot  table [x=xValues, y=sch_naturalScheduling, col sep=comma] {\folderName/schPerformanceForEqualPowerF.csv};
\addplot  table [x=xValues, y=sch_optimizedBlockInterleaver1, col sep=comma] {\folderName/schPerformanceForEqualPowerF.csv};
\addplot  table [x=xValues, y=sch_Peng1, col sep=comma] {\folderName/schPerformanceForEqualPowerF.csv};
\addplot  table [x=xValues, y=sch_Heuristic1, col sep=comma] {\folderName/schPerformanceForEqualPowerF.csv};
\addplot  table [x=xValues, y=sch_gurobi, col sep=comma] {\folderName/schPerformanceForEqualPowerF.csv};
\coordinate (top) at (rel axis cs:0,1);

\nextgroupplot[title=$w$, title style={yshift=2ex,},
xmin = 3,
xmax = 25,
ymax = \ymaxBandC, 
ytick =  \yTicksBandC,
xtick =  {5,10,15,20,25},
xlabel=$N$,
extra x ticks =  {3,4,5,6,7,8,9,10,11,12,13,14,15,16,17,18,19,20,21,22,23,24,25},
every extra x tick/.style={
	xtick pos=right,
	xticklabel pos=right,		
	xticklabels from table={\datatableN}{W},
}
]
\addplot  table [x=xValues, y=sch_naturalScheduling, col sep=comma] {\folderName/schPerformanceForEqualPowerN.csv};
\addplot  table [x=xValues, y=sch_optimizedBlockInterleaver1, col sep=comma] {\folderName/schPerformanceForEqualPowerN.csv};
\addplot  table [x=xValues, y=sch_Peng1, col sep=comma] {\folderName/schPerformanceForEqualPowerN.csv};
\addplot  table [x=xValues, y=sch_Heuristic1, col sep=comma] {\folderName/schPerformanceForEqualPowerN.csv};
\addplot  table [x=xValues, y=sch_gurobi, col sep=comma] {\folderName/schPerformanceForEqualPowerN.csv};

\nextgroupplot[
xmin = 1,
xmax = 10,
ymax = \ymaxA,
xlabel=$T$,
xtick =  \plotAxTicklabels,
]
\addplot +[restrict expr to domain={\coordindex}{0:9}]   table [x=xValues, y=sch_naturalScheduling, col sep=comma] {\folderName/schPerformanceForHeuristic1PowerT.csv};
\addplot +[restrict expr to domain={\coordindex}{0:9}]   table [x=xValues, y=sch_optimizedBlockInterleaver1, col sep=comma] {\folderName/schPerformanceForHeuristic1PowerT.csv};
\addplot +[restrict expr to domain={\coordindex}{0:9}]   table [x=xValues, y=sch_Peng1, col sep=comma] {\folderName/schPerformanceForHeuristic1PowerT.csv};
\addplot +[restrict expr to domain={\coordindex}{0:9}]   table [x=xValues, y=sch_Heuristic1, col sep=comma] {\folderName/schPerformanceForHeuristic1PowerT.csv};
\addplot +[restrict expr to domain={\coordindex}{0:9}]   table [x=xValues, y=sch_gurobi, col sep=comma] {\folderName/schPerformanceForHeuristic1PowerT.csv};

\nextgroupplot[
xmin = 6,
xmax = 20,
ymax = \ymaxBandC, ymin = 0.98,
ytick =  \yTicksBandC,
xlabel=$F$,
xtick =  {6,12,20},
]
\addplot  table [x=xValues, y=sch_naturalScheduling, col sep=comma] {\folderName/schPerformanceForHeuristic1PowerF.csv};
\addplot  table [x=xValues, y=sch_optimizedBlockInterleaver1, col sep=comma] {\folderName/schPerformanceForHeuristic1PowerF.csv};
\addplot  table [x=xValues, y=sch_Peng1, col sep=comma] {\folderName/schPerformanceForHeuristic1PowerF.csv};
\addplot  table [x=xValues, y=sch_Heuristic1, col sep=comma] {\folderName/schPerformanceForHeuristic1PowerF.csv}; 
\addplot  table [x=xValues, y=sch_gurobi, col sep=comma] {\folderName/schPerformanceForHeuristic1PowerF.csv};

\nextgroupplot[
xmin = 3,
xmax = 25,
ymax = \ymaxBandC, 
ytick =  \yTicksBandC,
xtick =  {5,10,15,20,25},
xminorgrids,
xlabel=$N$,
]
\addplot  table [x=xValues, y=sch_naturalScheduling, col sep=comma] {\folderName/schPerformanceForHeuristic1PowerN.csv};
\addplot  table [x=xValues, y=sch_optimizedBlockInterleaver1, col sep=comma] {\folderName/schPerformanceForHeuristic1PowerN.csv};
\addplot  table [x=xValues, y=sch_Peng1, col sep=comma] {\folderName/schPerformanceForHeuristic1PowerN.csv};
\addplot  table [x=xValues, y=sch_Heuristic1, col sep=comma] {\folderName/schPerformanceForHeuristic1PowerN.csv};
\addplot  table [x=xValues, y=sch_gurobi, col sep=comma] {\folderName/schPerformanceForHeuristic1PowerN.csv};

\nextgroupplot[
xmin = 1,
xmax = 10,
ymax = \ymaxA,
xlabel=$T$,
xtick =  \plotAxTicklabels,
]
\addplot +[restrict expr to domain={\coordindex}{0:9}]   table [x=xValues, y=sch_naturalScheduling, col sep=comma] {\folderName/schPerformanceForGurobiPowerT.csv};
\addplot +[restrict expr to domain={\coordindex}{0:9}]   table [x=xValues, y=sch_optimizedBlockInterleaver1, col sep=comma] {\folderName/schPerformanceForGurobiPowerT.csv};
\addplot +[restrict expr to domain={\coordindex}{0:9}]   table [x=xValues, y=sch_Peng1, col sep=comma] {\folderName/schPerformanceForGurobiPowerT.csv};
\addplot +[restrict expr to domain={\coordindex}{0:9}]   table [x=xValues, y=sch_Heuristic1, col sep=comma] {\folderName/schPerformanceForGurobiPowerT.csv};
\addplot +[restrict expr to domain={\coordindex}{0:9}]   table [x=xValues, y=sch_gurobi, col sep=comma] {\folderName/schPerformanceForGurobiPowerT.csv};

\nextgroupplot[
xmin = 6,
xmax = 20,
ymax = \ymaxBandC, ymin = 0.98,
ytick =  \yTicksBandC,
xlabel=$F$,
xtick =  {6,12,20},
] 
\addplot  table [x=xValues, y=sch_naturalScheduling, col sep=comma] {\folderName/schPerformanceForGurobiPowerF.csv};
\addplot  table [x=xValues, y=sch_optimizedBlockInterleaver1, col sep=comma] {\folderName/schPerformanceForGurobiPowerF.csv};
\addplot  table [x=xValues, y=sch_Peng1, col sep=comma] {\folderName/schPerformanceForGurobiPowerF.csv};
\addplot  table [x=xValues, y=sch_Heuristic1, col sep=comma] {\folderName/schPerformanceForGurobiPowerF.csv}; 
\addplot  table [x=xValues, y=sch_gurobi, col sep=comma] {\folderName/schPerformanceForGurobiPowerF.csv};

\nextgroupplot[
xmin = 3,
xmax = 25,
ymax = \ymaxBandC, 
ytick =  \yTicksBandC,
xtick =  {5,10,15,20,25},
xminorgrids,
xlabel=$N$,
]
\addplot  table [x=xValues, y=sch_naturalScheduling, col sep=comma] {\folderName/schPerformanceForGurobiPowerN.csv};
\addplot  table [x=xValues, y=sch_optimizedBlockInterleaver1, col sep=comma] {\folderName/schPerformanceForGurobiPowerN.csv};
\addplot  table [x=xValues, y=sch_Peng1, col sep=comma] {\folderName/schPerformanceForGurobiPowerN.csv};
\addplot  table [x=xValues, y=sch_Heuristic1, col sep=comma] {\folderName/schPerformanceForGurobiPowerN.csv};
\addplot  table [x=xValues, y=sch_gurobi, col sep=comma] {\folderName/schPerformanceForGurobiPowerN.csv};

\coordinate[right=-5.7cm] (bot) at (rel axis cs:1,0);
\end{groupplot}
\node[below = 1cm of my plots c1r1.south] {(a) Equal Power ($F=20, N=20$)};     
\node[below = 1cm of my plots c2r1.south] {(b) Equal Power ($T=1, N=20$)};
\node[below = 1cm of my plots c3r1.south] {(c) Equal Power ($F=20,  T=1$)};

\node[below = 1cm of my plots c1r2.south] {(d) Heuristic Power  ($F=20, N=20$)};
\node[below = 1cm of my plots c2r2.south] {(e) Heuristic Power  ($T=1, N=20$)};
\node[below = 1cm of my plots c3r2.south] {(f) Heuristic Power  ($F=20,  T=1$)};

\node[below = 1cm of my plots c1r3.south,text width=4.5cm,align=center] {(g) Optimal power (numerical) \\($F=20, N=20$)};
\node[below = 1cm of my plots c2r3.south,text width=4.5cm,align=center] {(h) Optimal power (numerical) \\($T=1, N=20$)};
\node[below = 1cm of my plots c3r3.south,text width=4.5cm,align=center] {(i) Optimal power (numerical) \\($F=20,  T=1$)};

\path (top-|current bounding box.west)--
node[anchor=south,rotate=90] {Average number of successful links per VUE ($\bar{Z}$)}
(bot-|current bounding box.west);
\path[yshift=2cm] (top|-current bounding box.north)--
coordinate(legendpos)
(bot|-current bounding box.north);
\matrix[
matrix of nodes,
anchor=south,
draw,
inner sep=0.2em,
draw
]at([yshift=1ex]legendpos)
{
	\ref{plotsA:plot1}& BIS ($w=1$) &[5pt]
	\ref{plotsA:plot2}& BIS (optimized $w$) &[5pt]
	\ref{plotsA:plot2b}& \cite{Peng1} &[5pt]
	\ref{plotsA:plot3}& Heuristic scheduling &[5pt]
	\ref{plotsA:plot4}& Optimal scheduling (numerical) \\};
\end{tikzpicture}

\caption{\reviewerA{Average number of successful links for a VUE ($\bar{Z}$) for various scheduling algorithms}} \label{resultNSuccessfulLinks}
\end{figure*}

\endgroup


%% file: sec05_fairness_figureC.tex

\pgfplotscreateplotcyclelist{colorList1}{
	{black,mark=square},
	{blue,mark=triangle},
	{violet,mark=+},
	{green,mark=star},
	{red,mark=diamond},
}

\begingroup
\thickmuskip=0mu		

\begin{figure*}[!ht]
	\centering	

        \begin{tikzpicture}
\begin{groupplot}[cycle list name=colorList1, xmajorgrids, ymajorgrids,
	group style={group name=my plots,group size= 2 by 1,vertical sep=2.5cm ,horizontal sep=2cm },
	height=5.5cm,width=0.4\paperwidth	
	]

\nextgroupplot[
xmin = 0,
xmax = 9,
xlabel=\parbox{4cm}{\centering Number of successful links \\ for a VUE ($Z_i$) },
ylabel=CDF,
xtick =  {0,1,2,3,4,5,6,7,8,9},
ymax = 1,
]
\addplot table [x=xValues, y=sch_naturalScheduling, col sep=comma] {\folderName/cdfSchPerformanceForEqualPower.csv}; \label{plotsC:plot1} 
\addplot table [x=xValues, y=sch_optimizedBlockInterleaver1, col sep=comma] {\folderName/cdfSchPerformanceForEqualPower.csv}; \label{plotsC:plot2} 
\addplot table [x=xValues, y=sch_Peng1, col sep=comma] {\folderName/cdfSchPerformanceForEqualPower.csv}; \label{plotsC:plot2b} 
\addplot table [x=xValues, y=sch_Heuristic1, col sep=comma] {\folderName/cdfSchPerformanceForEqualPower.csv}; \label{plotsC:plot3} 
\addplot table [x=xValues, y=sch_gurobi, col sep=comma] {\folderName/cdfSchPerformanceForEqualPower.csv}; \label{plotsC:plot4} 

\nextgroupplot[
xmin = 1,
xmax = 20,
xlabel=VUE index ($i$),
ylabel=\parbox{4cm}{\centering Average number of successful links for a VUE ($\bar{Z}_i$) },
xtick =  {1,5,10,15,20},
ymax = 5,  
legend style={at={(-0.15,1.08)},anchor=south, legend columns=-1, column sep = 0.1cm}, 
]
\addplot table [x=xValues, y=sch_naturalScheduling, col sep=comma] {\folderName/eachVUESchPerformanceForEqualPower.csv}; \addlegendentry{\hspace{0.2cm} BIS ($w=1$) \em}
\addplot table [x=xValues, y=sch_optimizedBlockInterleaver1, col sep=comma] {\folderName/eachVUESchPerformanceForEqualPower.csv}; \addlegendentry{\hspace{0.2cm}BIS (optimized $w$) \em}
\addplot table [x=xValues, y=sch_Peng1, col sep=comma] {\folderName/eachVUESchPerformanceForEqualPower.csv}; \addlegendentry{\hspace{0.2cm} \cite{Peng1} \em}
\addplot table [x=xValues, y=sch_Heuristic1, col sep=comma] {\folderName/eachVUESchPerformanceForEqualPower.csv}; \addlegendentry{\hspace{0.2cm}Heuristic scheduling \em}
\addplot table [x=xValues, y=sch_gurobi, col sep=comma] {\folderName/eachVUESchPerformanceForEqualPower.csv}; \addlegendentry{\hspace{0.2cm} Optimal scheduling (numerical) \em}

%

\coordinate (bot) at (rel axis cs:1,0);
\end{groupplot}
\node[below = 1.3cm of my plots c1r1.south] {(a)};
\node[below = 1.3cm of my plots c2r1.south] {(b)};

\end{tikzpicture}

\caption{\reviewerA{Fairness comparison of number of successful links for equal power ($F=20,\, T=2,\, N=20$)}} \label{resultFairness}
\end{figure*}
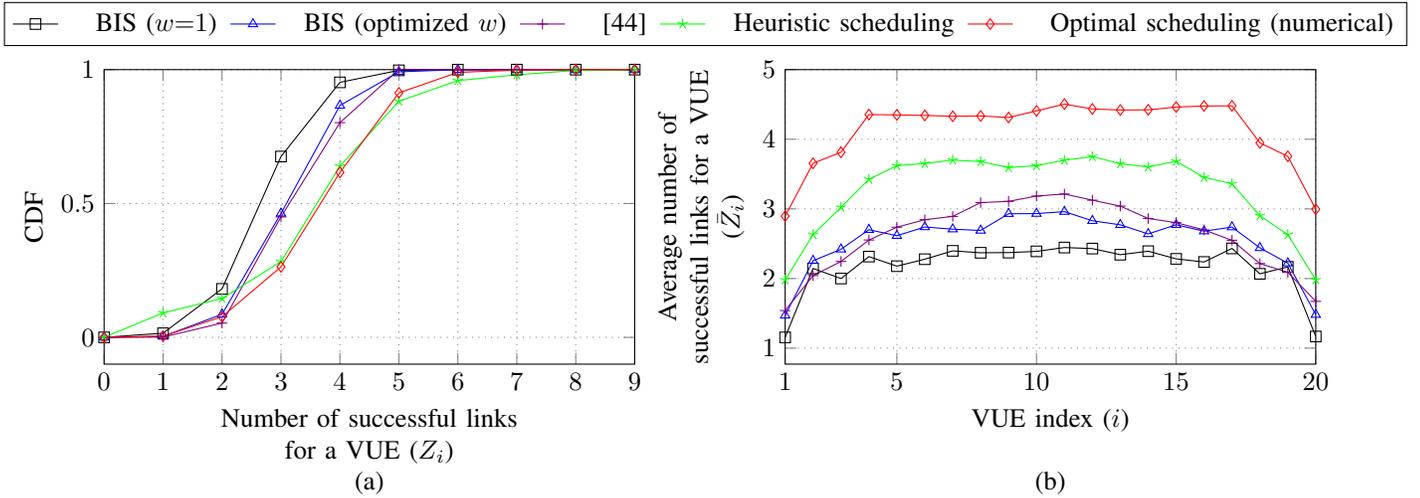

\endgroup


%% file: sec05_SimulationResults_figureB.tex

\pgfplotscreateplotcyclelist{colorList1}{
	{black,mark=square,mark options={fill=none}},
	{green,mark=star},
	{red,mark=diamond},
}

\begingroup
\thickmuskip=0mu		

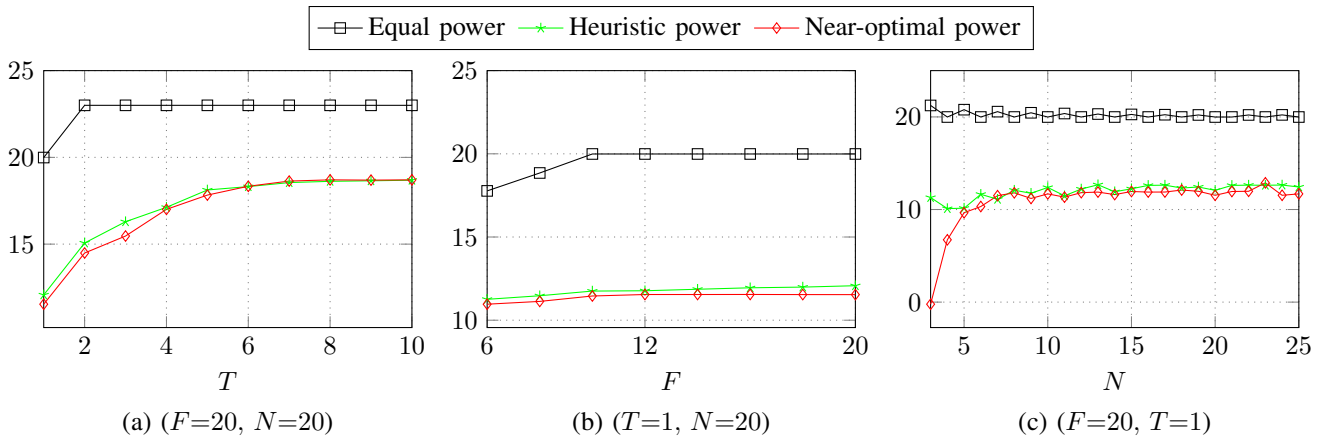
\begin{figure*}[!ht]
	\centering	

        \begin{tikzpicture}
\begin{groupplot}[cycle list name=colorList1, xmajorgrids, ymajorgrids,
	group style={group name=my plots,group size= 3 by 1,vertical sep=2.5cm },
	height=5cm,width=0.3\paperwidth	
	]

\nextgroupplot[
xmin=1, xmax=10,
xlabel=$T$,
xtick =  \plotAxTicklabels,
ymax = 25,
]
\addplot +[restrict expr to domain={\coordindex}{0:9}]   table [x=xValues, y=powerControl_equalPower, col sep=comma] {\folderName/powerValuesForNaturalScheduling_T.csv}; \label{plotsB:plot1} 
\addplot +[restrict expr to domain={\coordindex}{0:9}]   table [x=xValues, y=powerControl_kangWang, col sep=comma] {\folderName/powerValuesForNaturalScheduling_T.csv}; \label{plotsB:plot2} 
\addplot +[restrict expr to domain={\coordindex}{0:9}]   table [x=xValues, y=powerControl_gurobi, col sep=comma] {\folderName/powerValuesForNaturalScheduling_T.csv}; \label{plotsB:plot3} 

\nextgroupplot[
xmin=6, xmax=20,
xlabel=$F$,
xtick =  {6,12,20},
ymax = 25,  
legend style={at={(0.5,1.08)},anchor=south, legend columns=-1, column sep = 0.1cm}, 
]
\addplot  table [x=xValues, y=powerControl_equalPower, col sep=comma] {\folderName/powerValuesForNaturalScheduling_F.csv}; \addlegendentry{\hspace{-0.1cm}Equal power \em}
\addplot  table [x=xValues, y=powerControl_kangWang, col sep=comma] {\folderName/powerValuesForNaturalScheduling_F.csv};\addlegendentry{\hspace{-0.1cm}Heuristic power \em}
\addplot  table [x=xValues, y=powerControl_gurobi, col sep=comma] {\folderName/powerValuesForNaturalScheduling_F.csv};\addlegendentry{\hspace{-0.1cm}Near-optimal power \em}

\nextgroupplot[
xmin=3, xmax=25,
xlabel=$N$,
ymax = 25,
]
\addplot  table [x=xValues, y=powerControl_equalPower, col sep=comma] {\folderName/powerValuesForNaturalScheduling_N.csv};
\addplot  table [x=xValues, y=powerControl_kangWang, col sep=comma] {\folderName/powerValuesForNaturalScheduling_N.csv};
\addplot  table [x=xValues, y=powerControl_gurobi, col sep=comma] {\folderName/powerValuesForNaturalScheduling_N.csv};

\coordinate (bot) at (rel axis cs:1,0);
\end{groupplot}
\node[below = 1cm of my plots c1r1.south] {(a) ($F=20,\, N=20$)};
\node[below = 1cm of my plots c2r1.south] {(b) ($T=1,\, N=20$)};
\node[below = 1cm of my plots c3r1.south] {(c) ($F=20, \, T=1$)};

\end{tikzpicture}

\caption{Average transmit power per VUE (dBm) for various power control algorithms for BIS ($w=1$)}	 \label{resultTxPower}
\end{figure*}

\endgroup


%% file: sec06_Conclusion.tex
\section{Conclusions}  \label{sec:Conclusions}

This paper studies performance of V2V all-to-all broadcast communication by focusing more upon the scenario where CCI is limited due to the non-overlapping scheduling of VUEs. From the results presented in this paper, which are
for half-duplex communication, we can draw the following conclusions.
\begin{enumerate}
	\item Performance is mainly limited by ACI due to near-far situation in V2V networks when VUEs are multiplexed in frequency.
	\item Performance is heavily dependent on scheduling and power allocation.
	\item In general, scheduling with fixed
	and equal transmit powers is more effective in improving performance
	than subsequent power control.
	\item To find a schedule and power allocation to maximize
	performance can be stated as the nonconvex mixed integer quadratic constrained programming
	(MIQCP) problem in~(\ref{blp}).
	\item To find a schedule to maximize performance for a
	fixed power allocation can be stated as a Boolean linear
	programming (BLP) problem found by fixing $\Pb$ to a constant matrix in~(\ref{blp}). \label{alg:schedule}
	\item The heuristic scheduling algorithm for a fixed power allocation
	defined in Algorithm~\ref{heuristic1Algorithm} has significantly lower complexity than the BLP
	program and performs significantly better
	than the baseline block-interleaver scheduler defined in Algorithm~\ref{Alg:BIS}.
	\item To find a power allocation to maximize performance
	for a fixed schedule can be stated as the mixed integer linear
	programming (MILP) problem
	found by replacing the objective in~(\ref{blp}) with~\eqref{nearOptimalPC} and fixing
	$\mathbf{X}$. \label{alg:power}
	\item The heuristic power allocation algorithm for a fixed schedule
	defined in Algorithm~\ref{heuristicPowerControlAlgorithm} achieve similar performance as the solution to the MILP problem, but
	at a significantly lower computational complexity.
	items~\ref{alg:schedule}) and \ref{alg:power}) above, respectively.
\end{enumerate}

\section{Future Works}    \label{sec:future works}

\reviewerA{We note that the scalability is an issue for all the algorithms presented in this paper, since a centralized controller may not exist for a larger network and computing optimal solution becomes hard. One possible approach to reduce the computational complexity is to split the network into smaller networks and do the scheduling and power control for each smaller network separately. The splitting should be done in a ``soft'' manner to avoid the edge effects. For example, suppose $N$ VUEs are divided into $M$ groups and that each group has a centralized controller. We assume that the grouping is done such that VUEs in group $m$ want to communicate with VUEs found in groups $m-1$, $m$ and $m+1$ and that transmissions from group $m$ cause relative little interference to VUEs in groups $m\pm 2, m\pm 3, \ldots$. We partition the groups into 3 partitions, i.e., the groups $\{1,4,7,\ldots\}$ is called partition~1, groups $\{2,5,8,\ldots\}$ as partition~2, and groups $\{3,6,9,\ldots\}$ as partition~3. Since interference is limited between the groups within a partition, groups in each partition can reuse resources, e.g., groups $\{1,4,7,\ldots\}$ can reuse the same timeslot. However, since there can be interference between partitions, we use time-division multiplexing to separate partitions, e.g., the VUEs in partition 1 are scheduled in timeslots $\{1,3,5,\ldots\}$, partition~2 VUEs in timeslots $\{2,4,6,\ldots\}$, etc. In this way, there will be no inter-partition interference (CCI or ACI). The analysis of this scheme is not done yet, but will be presented in a future publication.}

\reviewerB{Additional future works would involve devising scheduling and power control algorithms for V2V communication networks in a decentralized manner (i.e., without a centralized controller), and to address the numerical sensitivity issues. A study upon the sensitivity of the parameters and the possibilities for multihop communication are also topics for future work.}


\begin{appendices} 

\section{Joint Scheduling and Power Control Problem Formulation by Focusing on Transmitter-Receiver Links}
\label{Appendix:powerControlFormulation}

Let us define $\mathbf{\Upsilon} \in \mathbb{R}^{N \times N \times T}$ with $\Upsilon_{i,j,t}$ being the SINR during timeslot $t$ for the link from VUE $i$ to VUE $j$, i.e., transmitter-receiver link $(i,j)$. The value of $\Upsilon_{i,j,t}$ can be computed as follows,
\begin{equation}
\Upsilon_{i,j,t} = \frac{ \overset{F}{\underset{f=1}{\sum}}   X_{i,f,t} P_{i,t} H_{i,j}} {\sigma^2 +  \sum\limits_{f=1}^{F} \overset{F}{\underset{\substack{f' = 1}}{\sum}}  \overset{N}{\underset{\substack{k=1 \\ k \neq i}}{\sum}}  X_{i,f,t} \acirf X_{k,f',t} P_{k,t} H_{k,j} } \label{Gammaijt_computation}
\end{equation}
where $\sigma^2$ is the noise variance and $P_{i,t}$ is the transmit power of VUE $i$ during timeslot $t$. 

Now we explain each component of (\ref{Gammaijt_computation}). Observe that $X_{i,f,t} P_{i,t} H_{i,j}$ in the numerator is the received signal power for the link $(i,j)$ on RB $(f,t)$, therefore, $\sum_f X_{i,f,t} P_{i,t} H_{i,j}$ is the total received signal power in timeslot $t$. Similarly  $\acirf X_{k,f',t} P_{k,t} H_{k,j}$ is the interference power received by VUE $j$ on RB $(f,t)$ from VUE $k$ when VUE $k$ is scheduled to transmit on RB $(f',t)$. Similarly, $X_{i,f,t} \acirf X_{k,f',t} P_{k,t} H_{k,j}$ is the same received interference power if VUE $i$ is scheduled to transmit in RB $(f,t)$. Therefore, $\sum_f \sum_{f'} \sum_{k \neq i} X_{i,f,t} \acirf X_{k,f',t} P_{k,t} H_{k,j}$ is the total interference power received to the link $(i,j)$ if VUE $i$ is scheduled to transmit in any of the RBs in timeslot $t$.

However, translating the constraint for achieving SINR target, i.e., $\Upsilon_{i,j,t} \geq \gammaT$, we get the following constraint,
\begin{align}
\overset{F}{\underset{f=1}{\sum}}   X_{i,f,t} P_{i,t} H_{i,j}  \hspace*{4cm}& \nonumber\\
-  \gammaT \sum\limits_{f=1}^{F} \overset{F}{\underset{\substack{f' = 1}}{\sum}}  \overset{N}{\underset{\substack{k=1 \\ k \neq i}}{\sum}}  X_{i,f,t} \acirf X_{k,f',t} P_{k,t} H_{k,j}  &\geq    \gammaT  \sigma^2
\end{align}
Observe that the above constraint is more complicated than a quadratic constraint. Moreover, we can simplify the above constraint only upto a Boolean quadratic constraint for a scheduling problem, upon fixing the power values $P_{i,t}\,\forall\,i,t$.

\section{Proving the Nonconvexity of (\ref{blpConstr1})}   \label{Appendix-nonConvexity}
Let us represent (\ref{blpConstr1}) as follows,
\begin{equation}
G(\Pb,\Xb,\Yb) \leq 0
\end{equation}
where $G(\Pb,\Xb,\Yb) $ is defined as follows,
\begin{align}
&G(\Pb,\Xb,\Yb) = \nonumber \\
& \quad -\overset{N}{\underset{i=1}{\sum}}   X_{i,f,t} P_{i,t} H_{i,j}   + \gammaT  \overset{F}{\underset{\substack{f' = 1 \\ f' \neq f}}{\sum}}  \overset{N}{\underset{k=1}{\sum}}  \acirf X_{k,f',t} P_{k,t} H_{k,j} \hspace*{1.5cm}    \nonumber \\
& \quad \quad + \gammaT \sigma^2 - \gammaT(N\Pmax+\sigma^2) (1-Y_{j,f,t}) \label{definitiong}
\end{align}

We prove the nonconvexity of (\ref{blpConstr1}) by proving that $G(\Pb,\Xb,\Yb) $ is nonconvex. We prove this by proving that the Hessian matrix of $G(\Pb,\Xb,\Yb) $ is not positive semidefinite, with respect to the two variables $x=X_{1,f,t}$ and $y=P_{1,t}$. The Hessian matrix of $G(\Pb,\Xb,\Yb) $ with respect to $x$ and $y$ is as follows,
\begin{equation}
\triangledown^2 G = \begin{bmatrix}
\frac{\partial^2 G}{\partial^2 x} & \frac{\partial^2 G}{\partial y \partial x} \\
\frac{\partial^2 G}{\partial x \partial y}  & \frac{\partial^2 G}{\partial^2 y} 
\end{bmatrix}
\end{equation}
However, observe that $\frac{\partial^2 G}{\partial^2 x} = \frac{\partial^2 G}{\partial^2 y} = 0$, and $\frac{\partial^2 G}{\partial x \partial y}  = \frac{\partial^2 G}{\partial y \partial x}$ from (\ref{definitiong}). Therefore, the determinant of the above Hessian matrix is $\left| \triangledown^2 G \right| = - (\frac{\partial^2 G}{\partial x \partial y})^2 \leq 0$. Since $\frac{\partial^2 G}{\partial x \partial y} \neq 0$ for some $j,f,t$, the corresponding determinant of the Hessian matrix is negative. Hence the function $G(\Pb,\Xb,\Yb) $ is nonconvex. This concludes the proof.

\section{Proving the Convergence of Algorithm \ref{heuristicPowerControlAlgorithm}}   \label{Appendix:ProvingConvergence}
\begin{lemma}	 \label{lemma2}
The Algorithm \ref{heuristicPowerControlAlgorithm} is convergent.
\end{lemma}

\begin{IEEEproof}
Observe that the set $\Ls$ is nonincreasing on each iteration. When the termination condition (Algorithm \ref{heuristicPowerControlAlgorithm}, line \ref{alg4:termination condition}) is not satisfied, the set of broken links $\Bs$ is nonempty. This implies that, the counter $C_{i,j}$ is incremented for some $(i,j) \in \Ls$ in each iteration. Therefore, the maximum number of iterations possible before the set $\Ls$ becomes empty is $\Ct\left|\Ls\right|$. This concludes the proof.
\end{IEEEproof}

\end{appendices}


%% file: Joint_Scheduling_and_PowerControl.bbl
\begin{thebibliography}{10}
\providecommand{\url}[1]{#1}
\csname url@samestyle\endcsname
\providecommand{\newblock}{\relax}
\providecommand{\bibinfo}[2]{#2}
\providecommand{\BIBentrySTDinterwordspacing}{\spaceskip=0pt\relax}
\providecommand{\BIBentryALTinterwordstretchfactor}{4}
\providecommand{\BIBentryALTinterwordspacing}{\spaceskip=\fontdimen2\font plus
\BIBentryALTinterwordstretchfactor\fontdimen3\font minus
  \fontdimen4\font\relax}
\providecommand{\BIBforeignlanguage}[2]{{%
\expandafter\ifx\csname l@#1\endcsname\relax
\typeout{** WARNING: IEEEtran.bst: No hyphenation pattern has been}%
\typeout{** loaded for the language `#1'. Using the pattern for}%
\typeout{** the default language instead.}%
\else
\language=\csname l@#1\endcsname
\fi
#2}}
\providecommand{\BIBdecl}{\relax}
\BIBdecl

\bibitem{Araniti1}
G.~Araniti, C.~Campolo, M.~Condoluci, A.~Iera, and A.~Molinaro, ``{LTE} for
  vehicular networking: a survey,'' \emph{IEEE Communications Magazine},
  vol.~51, no.~5, pp. 148--157, May 2013.

\bibitem{Wanlu2016}
W.~Sun, E.~G. Str\"om, F.~Br\"annstr\"om, K.~C. Sou, and Y.~Sui, ``Radio
  resource management for {D2D}-based {V2V} communication,'' \emph{IEEE
  Transactions on Vehicular Technology}, vol.~65, no.~8, pp. 6636--6650, Aug
  2016.

\bibitem{Abbas2}
T.~Abbas, F.~Tufvesson, K.~Sj\"{o}berg, and J.~Karedal, ``A measurement based
  shadow fading model for vehicle-to-vehicle network simulations,''
  \emph{International Journal of Antennas and Propagation}, 05 2015.

\bibitem{Taimoor}
D.~Vlastaras, T.~Abbas, M.~Nilsson, R.~Whiton, M.~Olb{\"a}ck, and F.~Tufvesson,
  ``\BIBforeignlanguage{eng}{Impact of a truck as an obstacle on
  vehicle-to-vehicle communications in rural and highway scenarios},'' in
  \emph{\BIBforeignlanguage{eng}{Proc.~{IEEE} 6th International Symposium on
  Wireless Vehicular Communications}}, Vancouver, Canada, 2014.

\bibitem{Meireles1}
R.~Meireles, M.~Boban, P.~Steenkiste, O.~Tonguz, and J.~Barros, ``Experimental
  study on the impact of vehicular obstructions in {VANETs},'' in \emph{2010
  IEEE Vehicular Networking Conference}, Dec 2010, pp. 338--345.

\bibitem{He1}
R.~He, A.~F. Molisch, F.~Tufvesson, Z.~Zhong, B.~Ai, and T.~Zhang,
  ``Vehicle-to-vehicle propagation models with large vehicle obstructions,''
  \emph{IEEE Transactions on Intelligent Transportation Systems}, vol.~15,
  no.~5, pp. 2237--2248, Oct 2014.

\bibitem{Hong1}
X.~Hong, Z.~Chen, C.-X. Wang, S.~A. Vorobyov, and J.~S. Thompson, ``Cognitive
  radio networks: Interference cancellation and management techniques,''
  \emph{IEEE Vehicular Technology Magazine}, vol.~4, no.~4, pp. 76--84, Dec.
  2009.

\bibitem{v2vsch1}
W.~Li, X.~Ma, J.~Wu, K.~S. Trivedi, X.~L. Huang, and Q.~Liu, ``Analytical model
  and performance evaluation of long-term evolution for vehicle safety
  services,'' \emph{IEEE Transactions on Vehicular Technology}, vol.~66, no.~3,
  pp. 1926--1939, March 2017.

\bibitem{v2vsch2}
J.~Zhou, R.~Q. Hu, and Y.~Qian, ``Message scheduling and delivery with
  vehicular communication network infrastructure,'' in \emph{2013 IEEE Global
  Communications Conference (GLOBECOM)}, Atlanta, USA, Dec 2013, pp. 575--580.

\bibitem{v2vsch3}
F.~Zeng, R.~Zhang, X.~Cheng, and L.~Yang, ``Channel prediction based scheduling
  for data dissemination in {VANETs},'' \emph{IEEE Communications Letters},
  vol.~21, no.~6, pp. 1409--1412, 2017.

\bibitem{AnverICC}
A.~Hisham, W.~Sun, E.~G. Str\"om, and F.~Br\"annstr\"om, ``Power control for
  broadcast {V2V} communications with adjacent carrier interference effects,''
  in \emph{IEEE International Conference on Communications (ICC)}, Kuala
  Lumpur, Malaysia, May 2016.

\bibitem{Albasry1}
H.~Albasry, H.~Zhu, and J.~Wang, ``The impact of in-band emission interference
  in {D2D}-enabled cellular networks,'' in \emph{GLOBECOM 2017 - 2017 IEEE
  Global Communications Conference}, Dec 2017.

\bibitem{Li1}
D.~Li and Y.~Liu, ``In-band emission in {LTE-A D2D}: Impact and addressing
  schemes,'' in \emph{2015 IEEE 81st Vehicular Technology Conference (VTC
  Spring)}, May 2015.

\bibitem{Angelakis1}
V.~Angelakis, S.~Papadakis, V.~A. Siris, and A.~Traganitis, ``Adjacent channel
  interference in 802.11a is harmful: Testbed validation of a simple
  quantification model,'' \emph{IEEE Communications Magazine}, vol.~49, no.~3,
  pp. 160--166, March 2011.

\bibitem{aciw1}
L.~Wang, X.~Qi, and K.~Wu, ``Embracing adjacent channel interference in next
  generation {WiFi} networks,'' in \emph{2016 IEEE International Conference on
  Communications (ICC)}, Kuala Lumpur, Malaysia, May 2016.

\bibitem{aciw2}
A.~Adya, P.~Bahl, J.~Padhye, A.~Wolman, and L.~Zhou, ``A multi-radio
  unification protocol for {IEEE} 802.11 wireless networks,'' in \emph{First
  International Conference on Broadband Networks}, San Jose, CA, USA, Oct 2004,
  pp. 344--354.

\bibitem{aciw3}
J.~Nachtigall, A.~Zubow, and J.~P. Redlich, ``The impact of adjacent channel
  interference in multi-radio systems using {IEEE} 802.11,'' in \emph{2008
  International Wireless Communications and Mobile Computing Conference}, Crete
  Island, Greece, Aug 2008, pp. 874--881.

\bibitem{acib1}
W.~Li, J.~Chen, H.~Long, and B.~Wu, ``Performance and analysis on {LTE} system
  under adjacent channel interference of broadcasting system,'' in \emph{2012
  IEEE 12th International Conference on Computer and Information Technology},
  Chengdu, China, Oct 2012, pp. 290--294.

\bibitem{acib2}
Q.~Wang and X.~Li, ``Analysis of {LTE FDD and TD-LTE} combination network's
  interference,'' in \emph{2016 2nd IEEE International Conference on Computer
  and Communications (ICCC)}, Chengdu, China, Oct 2016, pp. 2332--2336.

\bibitem{acib3}
J.~Ribadeneira-Ramirez, G.~Martinez, D.~Gomez-Barquero, and N.~Cardona,
  ``Interference analysis between digital terrestrial television {(DTT)} and
  {4G LTE} mobile networks in the digital dividend bands,'' \emph{IEEE
  Transactions on Broadcasting}, vol.~62, no.~1, pp. 24--34, March 2016.

\bibitem{acib4}
``{CEPT Report 40}: Compatibility study for {LTE and WiMAX} operating within
  the bands 880-915 {MHz} / 925-960 {MHz} and 1710-1785 {MHz} / 1805-1880 {MHz}
  (900/1800 {MHz} bands),'' Tech. Rep., {CEPT Electronic Commun. Committee
  (ECC)}, Nov 2010.

\bibitem{Xia1}
K.~Xia, Y.~Wang, and D.~Zhang, ``Coexistence interference evaluation and
  analysis of {LTE} with {3D-MIMO} system,'' in \emph{2017 IEEE 28th Annual
  International Symposium on Personal, Indoor, and Mobile Radio Communications
  (PIMRC)}, Oct 2017, pp. 1--6.

\bibitem{Campolo1}
C.~Campolo, A.~Molinaro, and A.~Vinel, ``Understanding adjacent channel
  interference in multi-channel {VANET}s,'' in \emph{2014 IEEE Vehicular
  Networking Conference (VNC)}, Dec 2014, pp. 101--104.

\bibitem{Campolo2}
C.~Campolo, C.~Sommer, F.~Dressler, and A.~Molinaro, ``On the impact of
  adjacent channel interference in multi-channel {VANET}s,'' in \emph{2016 IEEE
  International Conference on Communications (ICC)}, May 2016.

\bibitem{36.211}
\BIBentryALTinterwordspacing
3GPP, ``{Evolved Universal Terrestrial Radio Access (E-UTRA); Physical channels
  and modulation},'' {3rd Generation Partnership Project (3GPP)}, TR {36.211},
  March 2017. [Online]. Available:
  \url{http://www.3gpp.org/ftp/Specs/archive/36\_series/36.211}
\BIBentrySTDinterwordspacing

\bibitem{tommyVTC}
T.~Svensson and T.~Eriksson, ``On power amplifier efficiency with modulated
  signals,'' in \emph{Proc.~{IEEE} Vehicular Technology Conference}, Ottawa,
  Canada, May 2010.

\bibitem{Joao1}
\BIBentryALTinterwordspacing
J.~Almeida, M.~Alam, J.~Ferreira, and A.~S. Oliveira, ``Mitigating adjacent
  channel interference in vehicular communication systems,'' \emph{Digital
  Communications and Networks}, vol.~2, no.~2, pp. 57 -- 64, 2016. [Online].
  Available:
  \url{http://www.sciencedirect.com/science/article/pii/S2352864816300104}
\BIBentrySTDinterwordspacing

\bibitem{36.942}
\BIBentryALTinterwordspacing
3GPP, ``{Evolved Universal Terrestrial Radio Access (E-UTRA); Radio Frequency
  (RF) system scenarios},'' {3rd Generation Partnership Project (3GPP)}, TR
  {36.942}, Oct. 2014. [Online]. Available:
  \url{http://www.3gpp.org/ftp/Specs/html-info/36942.htm}
\BIBentrySTDinterwordspacing

\bibitem{802.11p}
IEEE, ``{ISO/IEC/IEEE - International Standard - Information
  technology--Telecommunications and information exchange between
  systems--Local and metropolitan area networks--Specific requirements--Part
  11: Wireless LAN medium access control (MAC) and physical layer (PHY)
  specifications},'' \emph{ISO/IEC/IEEE 8802-11:2018(E)}, pp. 1--3538, May
  2018.

\bibitem{DahlmanParkvallSkold}
E.~Dahlman, S.~Parkvall, and J.~Sk\"old, \emph{{4G: LTE/LTE-Advanced} for
  Mobile Broadband}.\hskip 1em plus 0.5em minus 0.4em\relax Oxford: Academic
  Press, 2011.

\bibitem{Florian1}
A.~Florian, Potra, and J.~W. Stephen, ``Interior-point methods,'' \emph{Journal
  of Computational and Applied Mathematics}, vol. 124, no.~1, pp. 281--302,
  Dec. 2000.

\bibitem{KangWang}
K.~Wang, C.~Chiasserini, J.~Proakis, and R.~Rao, ``Joint scheduling and power
  control for multicasting in wireless ad hoc networks,'' in \emph{Proc.~{IEEE}
  Vehicular Technology Conference}, vol.~5, Florida, USA, Oct. 2003, pp.
  2915--2920.

\bibitem{Karedal}
J.~Karedal, N.~Czink, A.~Paier, F.~Tufvesson, and A.~Molisch, ``Path loss
  modeling for vehicle-to-vehicle communications,'' \emph{IEEE Transactions on
  Vehicular Technology}, vol.~60, no.~1, pp. 323--328, Jan. 2011.

\bibitem{Koufos1}
K.~Koufos and C.~P. Dettmann, ``Temporal correlation of interference in
  vehicular networks with shifted-exponential time headways,'' \emph{IEEE
  Wireless Communications Letters}, Aug. 2018.

\bibitem{Dewen1}
\BIBentryALTinterwordspacing
D.~Kong and X.~Guo, ``Analysis of vehicle headway distribution on multi-lane
  freeway considering car-truck interaction,'' \emph{Advances in Mechanical
  Engineering}, vol.~8, no.~4, 2016. [Online]. Available:
  \url{https://doi.org/10.1177/1687814016646673}
\BIBentrySTDinterwordspacing

\bibitem{Cowan1}
\BIBentryALTinterwordspacing
R.~J. Cowan, ``Useful headway models,'' \emph{Transportation Research}, vol.~9,
  no.~6, pp. 371 -- 375, 1975. [Online]. Available:
  \url{http://www.sciencedirect.com/science/article/pii/0041164775900088}
\BIBentrySTDinterwordspacing

\bibitem{Luttinen1}
\BIBentryALTinterwordspacing
R.~T. Luttinen, ``Statistical analysis of vehicle time headways,'' Ph.D.
  dissertation, 1996. [Online]. Available:
  \url{http://urn.fi/urn:nbn:fi:tkk-007970}
\BIBentrySTDinterwordspacing

\bibitem{36.885}
\BIBentryALTinterwordspacing
3GPP, ``{Technical Specification Group Radio Access Network; Study on
  {LTE-based V2X} Services},'' {3rd Generation Partnership Project (3GPP)}, TR
  {36.885}, June 2016. [Online]. Available:
  \url{http://www.3gpp.org/ftp/Specs/html-info/36885.htm}
\BIBentrySTDinterwordspacing

\bibitem{Abbas1}
T.~Abbas, J.~Nuckelt, T.~K\"urner, T.~Zemen, C.~F. Mecklenbr\"auker, and
  F.~Tufvesson, ``Simulation and measurement-based vehicle-to-vehicle channel
  characterization: Accuracy and constraint analysis,'' \emph{IEEE Transactions
  on Antennas and Propagation}, vol.~63, no.~7, pp. 3208--3218, July 2015.

\bibitem{Lin1}
L.~Cheng, B.~E. Henty, F.~Bai, and D.~D. Stancil, ``Highway and rural
  propagation channel modeling for vehicle-to-vehicle communications at 5.9
  {GHz},'' in \emph{2008 IEEE Antennas and Propagation Society International
  Symposium}, July 2008.

\bibitem{Kunisch1}
J.~Kunisch and J.~Pamp, ``Wideband car-to-car radio channel measurements and
  model at 5.9 {GHz},'' in \emph{Proceedings IEEE Vehicular Technology
  Conference}, Sept 2008.

\bibitem{36.300}
\BIBentryALTinterwordspacing
3GPP, ``{Evolved universal terrestrial radio access (E-UTRA) and evolved
  universal terrestrial radio access network (E-UTRAN); overall description;
  stage 2},'' {3rd Generation Partnership Project (3GPP)}, TR {36.300}, Mar.
  2017. [Online]. Available:
  \url{http://www.3gpp.org/ftp/Specs/html-info/36300.htm}
\BIBentrySTDinterwordspacing

\bibitem{AnverArchive}
\BIBentryALTinterwordspacing
A.~Hisham, E.~G. Str\"om, F.~Br\"annstr\"om, and L.~Yan, ``{Additional Results
  of Scheduling and Power Control for {V2V} Broadcast Communications with
  Co-Channel and Adjacent Channel Interference},'' Tech. Rep., Dec 2018.
  [Online]. Available:
  \url{https://arxiv.org/src/1708.02444/anc/Additional\_Results.pdf}
\BIBentrySTDinterwordspacing

\bibitem{Peng1}
B.~Peng, C.~Hu, T.~Peng, Y.~Yang, and W.~Wang, ``A resource allocation scheme
  for {D2D} multicast with {QoS} protection in {OFDMA}-based systems,'' in
  \emph{2013 IEEE 24th Annual International Symposium on Personal, Indoor, and
  Mobile Radio Communications (PIMRC)}, Sept 2013, pp. 2383--2387.

\bibitem{Gurobi}
\BIBentryALTinterwordspacing
{Gurobi Optimization, Inc.}, ``Gurobi optimizer reference manual,'' 2015.
  [Online]. Available: \url{http://www.gurobi.com}
\BIBentrySTDinterwordspacing

\bibitem{AnverMatlabCode}
\BIBentryALTinterwordspacing
\emph{Matlab code for {V2V} communication}, Oct. 2018. [Online]. Available:
  \url{https://github.com/anverhisham/Scheduling\_and\_powerControl\_ \\
  in\_V2V\_Communication}
\BIBentrySTDinterwordspacing

\end{thebibliography}
